\documentclass{aa}
\usepackage{txfonts}
\usepackage{psfig}
\usepackage{natbib}
\bibpunct{(}{)}{;}{a}{}{,}

\begin{document}

\def\eps{\varepsilon}
\def\aap{A\&A}
\def\apj{ApJ}
\def\aapr{A\&A Rev.}
\def\apjl{ApJL}
\def\mnras{MNRAS}
\def\araa{ARA\&A}
\def\aj{AJ}
\def\qjras{QJRAS}
\def\physrep{Phys. Rep.}
\def\nat{Nature}
\def\aaps{A\&A Supp.}
\def\lesssim{\mathrel{\hbox{\rlap{\hbox{\lower4pt\hbox{$\sim$}}}\hbox{$<$}}}}
\def\gtrsim{\mathrel{\hbox{\rlap{\hbox{\lower4pt\hbox{$\sim$}}}\hbox{$>$}}}}

\def\vx{\vec{x}}
\def\vr{\vec{r}}
\def\vxp{\vec{x}_\perp}
\def\vrp{\vec{r}_\perp}
\def\vk{\vec{k}}
\def\vkp{\vec{k}_\perp}

\def\RM{{\rm RM}}
\def\obs{{\rm obs}}
 
\def\C#1{#1}
\def\del#1{}

\title{The Magnetic Power Spectrum in Faraday Rotation Screens}
\titlerunning{Faraday Rotation Screens}
\author{Torsten A. En{\ss}lin \and Corina Vogt}
\authorrunning{T. A. En{\ss}lin \and C. Vogt}
\institute{Max-Planck-Institut f\"{u}r
Astrophysik, Karl-Schwarzschild-Str.1, Postfach 1317, 85741 Garching,
Germany} 
\date{Submitted 29.7.2002 / Accepted 7.2.2003 }

\abstract{
The autocorrelation function and similarly the Fourier-power spectrum
of a rotation measure (RM) map of an extended background radio source
can be used to measure components of the magnetic autocorrelation and
power-spectrum tensor within a foreground Faraday screen. It is
possible to reconstruct the full non-helical part of this tensor in
the case of an isotropic magnetic field distribution statistics. The
helical part is only accessible with additional information; e.g. the
knowledge that the fields are force-free. The magnetic field strength,
energy spectrum and autocorrelation length $\lambda_B$ can be obtained
from the non-helical part alone. We demonstrate that $\lambda_B$ can
differ substantially from $\lambda_\RM$, the observationally easily
accessible autocorrelation length of an RM map. In typical
astrophysical situation $\lambda_\RM > \lambda_B$. Any RM study, which does not take this distinction
into account, likely underestimates the magnetic field strength.
For power-law magnetic power spectra, and for patchy magnetic
field configurations the central RM autocorrelation function is shown
to have characteristic asymptotic shapes. Ways to constrain the volume
filling factor of a patchy field distribution are discussed.
We discuss strategies to analyse observational data, taking into
account -- with the help of a window function -- the limited extent of
the polarised radio source, the spatial distribution of the electron
density and average magnetic energy density in the screen, and
allowing for noise reducing data weighting. We briefly discuss the
effects of possible observational artefacts, and strategies to avoid
them.
 }\maketitle

\section{Introduction\label{sec:intro}}
\subsection{Cosmic magnetic fields}

The interstellar and intergalactic plasma is magnetised. The origin of
the magnetic fields is partly a mystery, yet it allows fascinating
insights into dynamical processes in the Universe.  Magnetic fields
are an important constituent of cosmic plasma in so far as they couple
the often collisionless charged particles by the Lorentz-force. They
are able to inhibit transport processes like heat conduction, spatial
mixing of gas, and propagation of cosmic rays.  They are essential for
the acceleration of cosmic rays.  They mediate forces through their
tension and pressure, giving the plasma additional macroscopic degrees
of freedom in terms of Alfv\'enic and magnetosonic waves. They allow
distant cosmic ray electron populations to be observed by
magneto-curvature (synchrotron) radiation.

Observational studies of spiral galaxies have revealed highly
organised magnetic field configurations, often in alignment with the
optical spiral arms. These magnetic fields are believed to be
generated and shaped by the dynamo action of the differentially
rotating galaxy disks from some initial weak seed fields.

The seed fields could have many origins, ranging from outflows from
stars and active galactic nuclei over battery effects in shock waves,
in ionisation fronts, and in neutral gas-plasma interactions, up to being
primordially generated in high energy processes like phase transitions
or inflation during the very early Universe.

In order to learn more about the magnetic field origin, less processed
plasma has to be studied. There is the possibility that the magnetic
fields outside galaxies, in galaxy clusters and -- if existing -- even
in the wider intergalactic space carry more information on the field's
origin. In clusters, magnetic fields with a much lower degree of
ordering, compared to the organised fields in spiral galaxies, have
been detected. However, they may be highly processed by turbulent gas
flows driven by galaxy cluster mergers, which may mask their
origin. Regardless, cluster magnetic fields are an interesting laboratory
to study magneto-hydrodynamical (MHD) turbulence, and are of great
importance to understand thermal and non-thermal phenomena in the
intra-cluster medium.

Despite their obvious importance for many astrophysical questions, and
despite many observational efforts to measure their properties, our
knowledge of galactic and intergalactic magnetic fields is still
poor. For an overview on the present observational and theoretical
knowledge the excellent review articles by \cite{1987QJRAS..28..197R},
\cite{1993A&ARv...4..449W}, \cite{1994RPPh...57..325K},
\cite{1996ARA&A..34..155B}, \cite{1999ARA&A..37...37K},
\cite{2001SSRv...99..243B}, \cite{2001PhR...348..163G},
\cite{2002ARA&A..40..319C}, and \cite{2002RvMP...74..775W} should be
consulted.

\subsection{Faraday rotation}

One way to probe magnetic fields is to use the Faraday rotation
effect. Linearly polarised radio emission experiences a rotation of
the polarisation plane when it transverses a plasma with a non-zero
magnetic field component along its propagation direction. If the
Faraday active medium is external to the source, a wavelength-square
dependence of the polarisation angle measured can be observed and used
to obtain the RM, which is the proportionality constant of this
dependence. Such situations are realised in nature in cases where a
polarised radio galaxy is located behind the magnetised medium of a
galaxy, or behind or embedded in a galaxy cluster.

The focus of this work is on the analysis of RM maps of Faraday
screens, in which the fields are statistically isotropically
distributed. This should be approximately fulfilled in galaxy
clusters, but not in the highly organised spiral galaxies. However,
our analysis should also give some insight into the statistics of RM
maps of galaxies, since many of the results do not strictly require
perfect isotropy.

Magnetic fields in galaxy clusters are known to exist due to detection
of cluster wide synchrotron emission \citep{1970MNRAS.151....1W}, and
detection of their Faraday rotation effect. Although the association
of the RM with the intra-cluster medium is not unambiguous, since it
could also be produced in a magnetised plasma skin of the observed
radio galaxy \citep{1990ApJ...357..373B}, there are arguments in
favour of such an interpretation: (i) The asymmetric depolarisation of
double radio lobes embedded in galaxy clusters can be understood as
resulting from a difference in the Faraday depth of the two lobes
\citep{1988Natur.331..149L,1988Natur.331..147G}. (ii) A recent RM
study by \cite{2001ApJ...547L.111C} of point sources located mostly
behind (but 40\% inside) galaxy clusters show a larger dispersion in
RM values than a reference sample without a galaxy cluster
intersecting the line-of-sight. (iii) The cluster-wide radio halos
observed in some clusters of galaxies
\cite[e.g.][]{1999dtrp.conf....3F} show synchrotron emission of
relativistic electrons within magnetic fields. The cluster fields
strength should be within an order of magnitude of their Faraday
rotation estimates for the radio emitting electrons to have a
reasonable energy density (compared to the thermal one) \cite[e.g. as
can be read off Fig.~1 in][]{1998AA...330...90E}.

Typical RM values of galaxy clusters are of the order of a few 100
rad/m$^2$, being consistent with field strengths of a few $\mu$G which
are well below equipartition with the thermal cluster gas. However, in
cooling flow clusters extreme RM values of a few 1000 rad/m$^2$ were
detected \cite[see][]{2002ARA&A..40..319C}, indicating possibly
substantial magnetic pressure support of the intra-cluster gas there.

Although the magnetic fields of galaxy clusters are less ordered than
these of spiral galaxies, the presence of coherent structures is
suggested by high resolution Faraday maps, which exhibit sometimes RM
bands \citep[e.g.][]{1987ApJ...316..611D, 1993ApJ...416..554T,
2001MNRAS.326....2T, 2002ApJ...567..202E}. Such bands may be caused by
shear-amplification of originally small-scale magnetic fields, as seen
in numerical MHD simulations of galaxy cluster formation
\citep{1999A&A...348..351D}. They are likely embedded within a
magnetic power-spectrum which extends over several orders of magnitude
in wavevector space. Similar to hydrodynamical turbulence, a broad
energy injection range is followed by a power-law spectrum at larger
wavevectors. For attempts to measure the magnetic power spectrum from
cluster simulations and radio maps see \cite{2002A&A...387..383D} and
\cite{astro-ph/0211292} respectively\footnote{The conventions to
describe the spectra may differ in these articles from the one used
here.}.

Another area of application of the theory developed here can be to
measure the properties of an hypothetical large-scale magnetic field
outside clusters of galaxies, which could be of primordial origin. A
pioneering feasibility study in this direction was done by
\cite{1998ApJ...495..564K}, who already outlined several of the ideas
investigated in this work. He proposed to probe the cosmological
magnetic fields by using catalogues of RM measurements of distant
radio galaxies and to measure spatial RM correlations between them in
order to measure the magnetic power spectrum, as we propose to do for
extended Faraday rotation maps. A RM search for fields in the
Lyman-$\alpha$ forest by \cite{1995ApJ...445..624O} found at most a
marginal detection. If they exist, primordial magnetic fields may be
detectable by Faraday rotation of the cosmic microwave background
(CMB) polarisation during and shortly after the epoch of
recombination, as \cite{1996ApJ...469....1K} proposed.

\cite{ohno2002} proposed to use the CMB polarisation even for RM
studies of nearby galaxy clusters.  Should this speculative proposal
become technically feasible, a lot of detailed information on
intra-cluster magnetic fields could be obtained.

\subsection{Philosophy of the paper}

If magnetic fields are sampled in a sufficiently large volume, they
can hopefully be regarded to be statistically homogeneous and
statistically isotropic. This means that any statistical average of a
quantity depending on the magnetic fields does not depend on the exact
location, shape, orientation and size of the used sampling volume.
The quantity we are interested in this paper is the autocorrelation
(or two-point-correlation) function (more exactly: tensor) of the
magnetic fields. The information contained in the autocorrelation
function is equivalent to the information stored in the
power-spectrum, as stated by the Wiener-Khinchin Theorem (WKT). We
therefore present two equivalent approaches, one based in real space,
and one based in Fourier space. The advantage of this redundancy is
that some quantities are easier accessible in one, and others in the
other space. Further, this allows to crosscheck computer algorithms
based on this work by comparing results gained by the different
approaches.
The observable we can use to access the magnetic fields is Faraday
rotation maps of extended polarised radio sources located behind a
Faraday screen.  Since an RM map shows basically the line-of-sight
projected magnetic field distribution, the RM autocorrelation function
is mainly given by the projected magnetic field autocorrelation function.
Therefore measuring the RM autocorrelation allows to measure the
magnetic autocorrelation, and thus provides a tool to estimate
magnetic field strength and correlation length. 
The situation is a bit more complicated than described above, due to
the vector nature of the magnetic fields. This implies that there is
an autocorrelation tensor instead of a function, which contains nine
numbers corresponding to the correlations of the different magnetic
components against each other, which in general can all be
different. The RM autocorrelation function contains only information
about one of these values, the autocorrelations of the magnetic field
component parallel to the line-of-sight. However, in many instances
the important symmetric part of the tensor can be reconstructed and
using this information the magnetic field strength and correlation
length can be obtained.  This is possible due to three observations:
\begin{enumerate}
\item
{\bf Magnetic isotropy}: If the sampling volume is sufficiently large,
so that the local anisotropic nature of magnetic field distributions
is averaged out, the (volume averaged) magnetic autocorrelation tensor
is isotropic. This means, that the diagonal elements of the tensor are
all the same, and that the off-diagonal elements are described by two
numbers, one giving their symmetric, and one giving their
anti-symmetric (helical) contribution.
\item
{\bf Divergence-freeness of magnetic fields}: The condition
$\vec{\nabla} \vec{\cdot} \vec{B} = 0$ ($\vec{B}$ is the magnetic
field) couples the diagonal and off-diagonal components of the
symmetric part of the autocorrelation tensor. Knowledge of one
diagonal element (e.g. from a RM measurement) therefore specifies
fully the symmetric part of the tensor. The trace of the
autocorrelation tensor, which can be called {\it scalar magnetic
autocorrelation function} $w(r)$, contains all the information
required to measure the average magnetic energy density $\eps_B =
w(0)/(8\,\pi)$ or the magnetic correlation length $\lambda_B = \int_{-\infty}^{\infty}
dr\,w(r)/w(0)$.
\item
{\bf Unimportance of helicity}: Although helicity is a crucial
quantity for the dynamics of magnetic fields, it does not enter
any estimate of the average magnetic energy density, or magnetic
correlation length, because helicity only affects off-diagonal terms
of the autocorrelation tensor. The named quantities depend only on the
trace of the tensor and are therefore unaffected by helicity. One
cannot measure helicity from a Faraday rotation map alone, since it
requires the comparison of two different components of the magnetic
fields, whereas the RM map contains information on only one component.
\end{enumerate}
In a realistic situation, the sampling volume is determined by the
shape of the polarised radio emitter and the geometry of the Faraday
screen, as given by the electron density and the magnetic field energy
density profile. The sampling volume can be described by a window
function, through which an underlying virtually statistical
homogeneous magnetic field is observed. The window function is zero
outside the probed volume, e.g. for locations which are not located in
front of the radio source. Inside the volume the window function
scales with the electron density (known from X-ray observations), with
the average magnetic energy profile (guessed from reasonable scaling
relations, but testable within the approach), and -- if wanted -- with
a noise reducing data weighting scheme. The effect of a finite window
function is to smear out the power in Fourier-space. Since this is an
unwanted effect one either has to find systems which provide a
sufficiently large window or one has to account for this bias. Since
the effect of a too small window on the results depends strongly on
the shape and size of the window at hand, a detailed discussion of all
the cases which can happen in practice is beyond the scope of this
paper. It has to be done for each application at hand separately.
Only idealised cases are discussed here for illustration. But
generally one can state, that the analysis is sensitive to magnetic
power on scales below a typical window size, and insensitive to scales
above.
The same magnetic power spectrum can have very different realizations,
since all the phase information is lost in measuring the power
spectrum \citep[for an instructive  visualisation of this
see][]{2001ApJ...554.1175M}. Since the presented approach relies on the
power spectrum only, it is not important if the magnetic fields are
highly organised in structures like flux-ropes, or magnetic sheets, or
if they are relatively featureless random-phase fields, as long as
their power spectrum is the same.
The autocorrelation analysis is fully applicable in all such
situations, as long as the fields are sampled with sufficient
statistics. The fact that this analysis is insensitive to
different realizations of the same power spectrum indicates that the
method is not able to extract all the information which may be in the
map. Additional information is stored in higher order correlation
functions, and such can in principle be used to make statements about
whether the fields are ordered or purely random (chaotic).  The
information on the magnetic field strength ($\vec{B}^2$, which is the
value at origin of the autocorrelation function), and correlation
length (an integral over the autocorrelation function) does only
depend on the autocorrelation function and not on the higher order
correlations.
The presented analysis relies on having a statistically isotropic
sample of magnetic fields, whereas MHD turbulence seems to be locally
inhomogeneous, which means that small scale fluctuations are
anisotropic with respect to the local mean field. However, whenever
the observing window is much larger than the correlation length of the
local mean field the autocorrelation tensor should be isotropic due to
averaging over an isotropic distribution of locally anisotropic
subvolumes. This works if not a preferred direction is superposed by
other physics, e.g. a galaxy cluster wide orientation of field lines
along a preferred axis. However, even this case can in principle be
treated by co-adding the RM signal from a sample of clusters, for
which a random distribution of such hypothetical axes can be
assumed. In any case, it is likely that magnetic anisotropy also
manifests itself in the Faraday rotation maps, since the projection
connecting magnetic field configurations and RM maps will conserve
anisotropy in most cases, except alignments by chance of the direction
of anisotropy and the line-of-sight. The presence of anisotropy can
therefore be tested, which is discussed later in great detail.
Since there are cases where already an inspection by eye seems to
reveal the existence of magnetic structures like flux ropes or
magnetic sheets, we briefly discuss their appearance in the
autocorrelation and the area filling statistics of RM maps. As already
stated, the presence of such structures does not limit our analysis,
as long as they are sufficiently sampled. Otherwise, one has to
replace e.g. the isotropy assumption by a suitable generalisation. In
many cases this will allow an analysis similar to the one proposed in
this paper. We leave this for future work and applications where this
might be required. A criteria to detect anisotropy statistically is
given in this work.

\subsection{Structure of the paper}

In Sect. \ref{sec:realspace} the autocorrelation functions of magnetic
fields and their RM maps are introduced, and their interrelation
investigated. In Sect. \ref{sec:Fspace} the same is done in Fourier
space, which has not only technical advantages, but also provides
insight into phenomena such as turbulence. Faraday map signatures of
magnetic structures like flux-ropes are briefly discussed in
Sect. \ref{sec:magStr}. Possible pitfalls due to observational
artefacts are investigated in Sect. \ref{sec:artefacts}. The
conclusions in Sect. \ref{sec:concl} summarise our main findings, and
give references to the important results and formulae in detail.

\section{Real space formulation\label{sec:realspace}}
\subsection{Basics\label{sec:rm}}

The Faraday rotation for a line-of-sight parallel to the $z$-axis and
displaced by $\vxp$ from it, which starts at a polarised radio source
at $z_{\rm s}(\vxp)$ and 
ends at the observer located at infinity is given by
\begin{equation}
\RM(\vxp) = a_0 \,\int_{z_{\rm s}(\vxp)}^\infty \!\!\!\!\!\!\! \!\!\!\! dz\; n_e(\vx) \,B_z(\vx)\,,
\end{equation}
where $a_0 = {e^3}/({2\pi \,m_e^2\,c^4})$, $\vx = (\vxp, z)$, $n_e$
the electron density, and $\vec{B}$ the magnetic field strength.  We
assume in the following that any Faraday rotation due to a foreground
as the Galaxy or the Earth's ionosphere is subtracted from the RM
values, and only the RM of the Faraday screen is remaining. We also
neglect any redshift effects, which can be included by inserting the
factor $(1+z_{\rm redshift}(z))^{-2}$ into the integrand.

The focus of this work is on the statistical expectation of the
two-point, or autocorrelation function of Faraday rotation maps. This
is defined by
\begin{equation}
C_\RM(\vxp,\vrp) = \langle \RM(\vxp) \,\RM(\vxp+\vrp) \rangle\,, 
\end{equation}
where the brackets indicate the expectation value of a statistical
ensemble average, which in practice may be replaced by a suitable
average, e.g. over an observed RM map. For a given polarised
background radio source of projected area $\Omega$ we define the
observable correlation function as
\begin{equation}
\label{eq:CRMobs}
C_{\rm RM}^\obs(\vrp) = \frac{1}{A_{\Omega}} \int
\!dx_\perp^2\, \RM(\vxp) \,\RM(\vxp+\vrp) \,, 
\end{equation}
where it is assumed that $\RM(\vxp) =0$ for $\vxp \notin
\Omega$. $A_{\Omega}$ is taken here to be the radio source area,
although other normalisations are imaginable, e.g. one might choose
the area over which $\vxp$ and $\vxp+\vrp$ are simultaneously within
$\Omega$. This latter weighting would give a statistically unbiased
estimator of $C_{\rm RM}(\vrp)$, but we do not recommend its usage for
the following reasons. The resulting estimator has a strong
sensitivity to poorly sampled variances on scales comparable to the
radio source diameter. On larger scales it is also ill-defined in the
mathematical sense. The normalisation proposed here leads to results
which do not suffer from this. It automatically down-weights the
signal from statistically insufficiently sampled baselines $\vrp$, and
this bias can be controlled (Sect. \ref{sec:windowcorr}). It further
turns out that the Fourier-space formulation of RM statistics
introduced in Sect. \ref{sec:Fspace} requires the weighting scheme
proposed here.

The RM signal from different subvolumes of the Faraday screen will
differ due to electron density and typical magnetic field strength
variations within the source. Such global variations can be regarded
as variations of a window function $f(\vx)$, which mediates the
relation between the observed RM signal and an underlying (rescaled)
magnetic field, which is virtually homogeneous in a statistical
sense. To be more specific, we choose a typical position $\vx_{\rm
ref}$ within the screen (e.g. the centre of a galaxy cluster), and define
$n_{\rm e0} = n_{\rm e}(\vx_{\rm ref})$ and $B_0 = \langle
B^2(\vx_{\rm ref}) \rangle^{1/2}$. We then define the window function
by
\begin{equation}
\label{eq:window}
f(\vx) = \vec{1}_{\{\vxp \in  \Omega\}}\,\vec{1}_{\{z\ge
z_{\rm s}(\vxp)\}}\, h(\vxp)\, g(\vx)\,{n_{\rm e}(\vx)}/{n_{\rm e0}}\,,
\end{equation}
where $\vec{1}_{\{condition\}}$ is defined to be $1$ if ${condition}$
is true, otherwise $0$. $g(\vx) = \langle B^2(\vx) \rangle^{1/2}/B_0$
is the dimensionless average magnetic field profile.  In galaxy
clusters a reasonable working assumption\footnote{We note that the
average magnetic field strength profile can be tested a posteriori for
consistency with the data, by comparing the observed values of
$\RM^2(\vx)$ to the model expectations (see Sect. \ref{sec:wtst}).}
for this may be $g(\vx) = (n_{\rm e}(\vx)/n_{\rm e0})^{\alpha_B}$, for
which we expect $\alpha_B \approx 1$
\cite[e.g. see][]{2001A&A...378..777D}.  The function $h(\vxp)$ allows
us to assign different pixels in the map different weights\footnote{If
$h(\vxp) \not = 1$ for some $\vxp\in \Omega$ the corresponding
weight has to be introduced into the analysis. The most efficient
way to do this is to make the data weighting virtually part of the
measurement process, by writing
$\RM(\vxp) \rightarrow \RM^{\rm w}(\vxp) = \RM(\vxp)\, h(\vxp)\,.$
For convenience, we drop the superscript$^{\rm w}$ again in the
following, and just note that the analysis described here has to be
applied to the weighted data $\RM^{\rm w}(\vxp)$.}
, e.g. in cases where the noise is a function of the position one
might want to down-weight noisy regions\footnote{If $\sigma(\vxp)$ is
the noise of the RM map at position $\vxp$ a reasonable choice of a
weighting function would be $h(\vxp) = \sigma_0/\sigma(\vxp)$. If the
noise map itself could have errors, the danger of over-weighting noisy
pixels with underestimated noise can be avoided by thresholding: If
$\sigma_0$ is a threshold below which the noise is regarded to be
tolerable, we propose to use $h(\vxp) = 1 /
(1+\sigma(\vxp)/\sigma_0)$. Another choice would be $h(\vxp) =
\vec{1}_{\{\sigma(\vxp) < \sigma_0\}}$ which just cuts out regions
which are recognised as too noisy.}. If no weighting applies
$h(\vxp) = 1$ everywhere.

The expectation of the observed RM correlations are
\begin{eqnarray}
\label{eq:exptCRMobs}
\langle C_{\rm RM}^\obs(\vrp) \rangle &=& \frac{a_0^2\,n_{\rm
e0}^2}{A_{\Omega}}  \int \!\!   d^3x  \int_{-\infty}^\infty \!\!\!\!\!\!\!
dr_z \nonumber\\ 
&& f(\vx) \, f(\vx+\vr) \, \langle \tilde{B}_z(\vx) 
\,\tilde{B}_z(\vx+\vr)\rangle,
\end{eqnarray}
with $\vr = (\vrp, r_z)$, and $\tilde{\vec{B}}(\vx) =
\vec{B}(\vx)/g(\vx)$ the rescaled magnetic field.  If properly
rescaled, the average strength of the field is independent of the
position. In that case, the rescaled magnetic field autocorrelation
tensor should also be independent of position:
\begin{equation}
\label{eq:defM}
M_{ij}(\vr) = \langle \tilde{B}_i(\vx)\,\tilde{B}_j(\vx + \vr)\rangle .
\end{equation}
If the spatial variation of the window function is on much larger
scales than the correlation length $\lambda_B$ of the magnetic fields,
then Eq.~\ref{eq:exptCRMobs} can be approximated to be
\begin{eqnarray}
\label{eq:CRM}
C_{\rm RM}(\vrp) \!&=&\! \langle C_{\rm RM}^\obs(\vrp) \rangle \!=\!
a_1\,C_\perp(\vrp),\,\mbox{with}\; a_1 \!=\! a_0^2\, n_{\rm e0}^2
\,L,\nonumber\\
\label{eq:Cperp1st}
C_\perp(\vrp) \!&=&\! \int_{-\infty}^{\infty}\!\! dr_z
M_{zz}(\vr),\; \mbox{and}\; \vr \!=\! (\vrp,r_z).
\end{eqnarray}
Here, we introduced the characteristic depth of the Faraday screen $L
= V_{[f]}/A_{\Omega}$, where $V_{[f]} =
\int\!dx^3\,f^2(\vx)$\label{sec:defVf} is the probed effective volume.
We also introduced for convenience the normalised RM autocorrelation
function $C_\perp$ which differs from $C_{\rm RM}$ only by a geometry
dependent factor $a_1$.

In the following we ignore the influence of the window function in the
discussion, since for sufficiently large windows it only affects
$a_1$. We therefore write $\vec{B}$ for $\tilde{\vec{B}}$ and keep in
mind that our measured field strength $B_0$ is estimated for a volume
close to the reference location $\vx_{\rm ref}$. At other locations,
the average magnetic energy density is given by $\eps_B(\vx) =
g^2(\vx) \, B_0^2$. This approach assumes implicitly that typical
length scales are the same throughout the Faraday screen. For
sufficiently extended screens, this assumption can be tested by
comparing results from different and separately analysed regions of
the RM map.

\subsection{Isotropic magnetic correlation tensor\label{sec:Mij}}

The magnetic autocorrelation tensor for homogeneous isotropic
turbulence, as assumed throughout the rest of this paper, can be
written as
\begin{equation}
\label{eq:Mreal}
M_{ij}(\vr) \!\!=\!\! M_{\rm N} (r)
\delta_{ij} + (M_{\rm L}(r)-M_{\rm N}(r))
\,\frac{r_i r_j}{r^2} + M_{\rm H}(r) \,\epsilon_{ijk} \,r_k
\end{equation}
\cite[e.g.][]{1999PhRvL..83.2957S} where the longitudinal, normal, and
helical autocorrelation functions, $M_{\rm L}(r)$, $M_{\rm N}(r)$, and
$M_{\rm H}(r)$ respectively, only depend on the distance, not on the
direction. The condition $\vec{\nabla} \vec{\cdot} \vec{B} = 0$ leads
to $\partial/\partial r_i\,M_{ij}(\vr) = 0$ (here and below we
make use of the sum convention). This allows us to connect the
non-helical correlation functions by
\begin{equation}
\label{eq:MLandMN}
M_{\rm N}(r) = \frac{1}{2 r} \frac{d}{d r} (r^2\, M_{\rm L}(r))
\end{equation}
\citep{1999PhRvL..83.2957S}.  The $zz$-component of the magnetic
autocorrelation tensor depends only on the longitudinal and normal
correlations, and not on the helical part:
\begin{equation}
M_{zz}(\vr) = M_{\rm L}(r)\,\frac{r^2_z}{r^2} + M_{\rm
N}(r)\,\frac{r^2_\perp}{r^2}\; \mbox{with}\; \vr = (\vrp, r_z)\, ,
\end{equation}
which implies that Faraday rotation is insensitive to magnetic
helicity. It is also useful to introduce the magnetic autocorrelation
function
\begin{equation}
\label{eq:def:w}
w(\vr) = \langle \vec{B}(\vx) \vec{\cdot} \vec{B}(\vx+\vr) \rangle = M_{ii}(\vr)\,,
\end{equation}
which is the trace of the autocorrelation tensor, and depends only on
$r$ (in the case of a statistically isotropic magnetic field
distribution, in the following called briefly {\it isotropic
turbulence}):
\begin{equation}
w(\vr) = w(r) = 2 M_{\rm N}(r) + M_{\rm L}(r) = \frac{1}{r^2} \frac{d}{dr}
(r^3\, M_{\rm L}(r))\,.
\end{equation}
In the last step Eq. \ref{eq:MLandMN} was used.  Since the average
magnetic energy density is given by $\langle \eps_B \rangle= w(0)/(8
\pi)$ the magnetic field strength can be determined by measuring the
zero-point of $w(r)$. This can be done by Faraday rotation
measurements: The RM autocorrelation can be written as
\begin{equation}
\label{eq:Cproj}
C_\perp(r_\perp) = \frac{1}{2} \int_{-\infty}^{\infty} \!\!\!\!\!\!
dr_z\, w(\sqrt{r_\perp^2 + r_z^2}) = \int_{r_\perp}^{\infty} \!\!\!\!\!\!
dr\, \frac{r\,w(r)}{\sqrt{r^2-r_\perp^2}}\,. 
\end{equation}
and is therefore just a line-of-sight projection of the magnetic
autocorrelations. Thus the magnetic autocorrelations $w(r)$ can be
derived from $C_\perp(r_\perp)$ by inverting an Abel integral
equation:
\begin{eqnarray}
\label{eq:Abel}
w(r) &=& -\frac{2}{\pi\,r}\,\frac{d}{dr}\,\int_r^{\infty} \!\!\!\!\!
dy\, \frac{y\, C_\perp(y)}{\sqrt{y^2 - r^2}}\\
\label{eq:Abel2}
&=& -\frac{2}{\pi}\,\int_r^{\infty} \!\!\!\!\!
dy\, \frac{C_\perp'(y)}{\sqrt{y^2 - r^2}}\,,
\end{eqnarray}
where the prime denotes a derivative. For the second equation it
was used that $w(r)$ stays bounded for $r \rightarrow \infty$.

Now, an observational program to measure magnetic fields is obvious:
From a high quality Faraday rotation map of a homogeneous, (hopefully)
isotropic medium of known geometry and electron density (e.g. derived
from X-ray maps) the RM autocorrelation has to be calculated
(Eq.~\ref{eq:CRM}). From this an Abel integration (Eq.~\ref{eq:Abel}
or \ref{eq:Abel2}) leads to the magnetic autocorrelation function,
which gives $\langle B^2\rangle$ at its origin. Formally,
\begin{equation}
\label{eq:B2first}
\langle B^2 \rangle = w(0) = -\frac{2}{\pi}\int_0^\infty \!\!\!\!\!\!  dy\,
\frac{C_\perp'(y)}{y}\,,
\end{equation}
but this formulation is notoriously sensitive to noise. More stable
methods are presented later.

\subsection{Correlation volume, area, and length}

For isotropic magnetic turbulence statements about integrals of $w(r)$
and $C_\perp(r_\perp)$ can be made. If correlations are short ranged,
in the sense that $M_{\rm L}(r)\,r^3 \rightarrow 0$ for $r \rightarrow
\infty$, e.g. because of a finite size of the magnetised volume, then
Eq.~\ref{eq:def:w} implies
\begin{equation}
\label{eq:ACvol}
V_B = \frac{4 \,\pi}{w(0)}\,\int_0^\infty \!\!\!\!\!\! dr\, r^2\, w(r) = 0\,.
\end{equation}
This means that the magnetic autocorrelation volume $V_B$ is zero and
$w(r)$ must therefore have positive and negative values if the
magnetic field is non-zero.  With the help of Eqs.  \ref{eq:Cproj} or
\ref{eq:Abel2} this can be translated into a statement on the more
directly measurable RM autocorrelation:
\begin{equation}
\label{eq:ACarea}
A_\RM = \frac{2 \,\pi}{C_\perp(0)}\, \int_0^\infty \!\!\!\!\!\! dr_\perp\, r_\perp\,
C_\perp(r_\perp) = 0\,.
\end{equation}
Also $C_\perp(r)$ must have positive and negative values for a
non-zero magnetic field, since the autocorrelation area $A_\RM$ of the
RM map is zero.  We note, that Eq.~\ref{eq:ACarea} can conveniently
be expressed as
\begin{equation}
\label{eq:ACarea2}
A_\RM = \frac{(\int\!\! dx_\perp \, \RM(\vxp))^2}{\int\!\! dx_\perp \,
\RM(\vxp)^2}\,,
\end{equation}
a form which should make clear that any foreground RM has to be
removed before this quantity can satisfy Eq.~\ref{eq:ACarea}.
Eqs.~\ref{eq:ACvol}, \ref{eq:ACarea}, and \ref{eq:ACarea2} are a
direct consequence of $\vec{\nabla} \vec{\cdot} \vec{B} = 0$.  These
equations can be used to test how much a given observation (or model)
deviates from giving a proper average of an isotropic ensemble.

For example, the frequently used {\it magnetic cell-model}, in which
cells of length-scale $l_{\rm cell}$ are filled with a from
cell-to-cell randomly oriented but internally homogeneous magnetic
field, does not fulfil $\vec{\nabla} \vec{\cdot} \vec{B} = 0$.
As a consequence it gives positive autocorrelation volumes and
surfaces of the order $V_B \sim l_{\rm cell}^3$ and $A_\RM \sim l_{\rm
cell}^2$ respectively. Therefore it should be possible to exclude such
an oversimplified model observationally.

Other integral quantities of the turbulent magnetic field to look at
are the existing non-zero magnetic and RM autocorrelation lengths.
The magnetic autocorrelation length can be defined as 
\begin{equation}
\label{eq:lB}
\lambda_B = \int_{-\infty}^\infty \!\!\!\!\!\! dr\, \frac{w(r)}{w(0)}
= 2 \frac{C_\perp(0)}{w(0)} \,,
\end{equation}
where for the derivation of the last expression Eq.~\ref{eq:Cproj} or
\ref{eq:Abel2} can be used. Even in globally homogeneous turbulence,
there is always a preferred direction defined by the local magnetic
field. One can ask for the correlation length along and perpendicular
to this locally defined direction and gets $\lambda_\| = \frac{3}{2}
\lambda_B,\;\mbox{and}\; \lambda_\perp = \frac{3}{4}
\lambda_B\;\mbox{so that}\; \lambda_B = \frac{1}{3}( \lambda_\| +
2\lambda_\perp ).$
\footnote{ The ratio of $\lambda_\|/\lambda_\perp = 2$ seems to be
small given the highly anisotropic nature of MHD turbulence
\cite[some recent progress on this can be found
in][]{1994ApJ...432..612S, 1997ApJ...485..680G, 2001ApJ...554.1175M,
2002ApJ...564..291C}. But it has to be noted, that there usually the
magnetic fluctuations on top of a mean field are investigated,
whereas here the correlation lengths of an isotropically oriented
magnetic field distribution sampled on length-scales above the
scale of the locally present mean field of these studies, is
investigated.
The ratio $\lambda_\|/\lambda_\perp = 2$ seems also to be in
contradiction to the typical MHD turbulence picture of long and thin
magnetic flux ropes with large aspect ratios. However, this is not
necessarily the case. For a given position $\vx_{\rm rope}$ on a flux
rope, the correlation length along and perpendicular will typically
have a much larger ratio than $\lambda_\|/\lambda_\perp = 2$. But the
position $\vx_{\rm rope}$ is a special position, since it was selected
to be a high field strength region, whereas the correlation length
$\lambda_\|$ and $\lambda_\perp$ are defined by the statistical
average over all positions in the volume. In order to make statements
about the presence or absence of magnetic filaments higher order
statistics, beyond the two-point level used mostly in this work, have
to be applied.  }

An observationally easily accessible length scale is the Faraday
rotation autocorrelation length:
\begin{equation}
\label{eq:lRM}
\lambda_{\rm RM} = \int_{-\infty}^{\infty} \!\!\!\!\!\! dr_\perp
\,\frac{C_\perp (r_{\perp})}{C_\perp (0)} =
\pi\,
\frac{\int_{-\infty}^{\infty}  dr \, r\,
w(r)}{\int_{-\infty}^{\infty}  dr \, w(r)}.
\end{equation}
From comparing Eqs.~\ref{eq:lB} and \ref{eq:lRM} it is obvious that
the RM correlation length-scale is not identical to the magnetic field
autocorrelation length. As shown later, the RM correlation length is
more strongly weighted towards the largest length-scales in the
magnetic fluctuation spectrum than the magnetic correlation
length. This is crucial, since in some cases these scales have been
assumed to be identical, which could have led to systematic
underestimates of magnetic field strengths.

Having now defined two characteristic length-scales of the fields and
their RM maps, suitable criteria testing the observed autocorrelation
volume and area for statistical completeness can be formulated:
$|V_B^\obs| \ll \lambda_B^3$ and $|A_\RM^\obs| \ll \lambda_\RM^2$.  We
note that for strictly positive and therefore unphysical
autocorrelation functions one would expect $|V_B^\obs| \sim
\lambda_B^3$ and $|A_\RM^\obs| \sim \lambda_\RM^2$ statistically.

This can be turned around. For a sufficiently large RM map of a
Faraday rotation screen with isotropic magnetic fields any homogeneous
RM contribution from additional magnetised foregrounds can relatively
accurately be measured. If there is a weak screen-intrinsic
homogeneous magnetic field component, its $z-$component is
\begin{equation}
\langle B_z \rangle = \frac{\int \!\! d^2x_\perp\, \RM(\vxp)}{a_0\, n_{\rm
e0}\, \int \!\! d^3x \, f(\vxp)}\,,
\end{equation}
where in $f(\vx)$ the average magnetic field profile may be set to
$g(\vx) =1$ everywhere for a truly homogeneous component.

\subsection{Testing the window function\label{sec:wtst}}

Now, all the necessary tools are introduced to test if the window
function $f(\vx)$ was based on a sensible model for the average
magnetic energy density profile $g^2(\vx)$ and the proper geometry of
the radio source within the Faraday screen $z_{\rm s}(\vxp)$ (see
Eq.~\ref{eq:window}).  Models can eventually be excluded a-posteriori
on the basis of
\begin{equation}
\label{eq:chi}
\chi^2(x_\perp) = \frac{\RM(\vxp)^2}{\langle \RM(\vxp)^2 \rangle}\,,
\end{equation}
where for the expected RM dispersion
\begin{equation}
\langle \RM(\vxp)^2 \rangle = \frac{1}{2}\, a_0^2\,n_{\rm 0}^2\,
B_0^2\, \lambda_B \int_{-\infty}^\infty \!\! \!\! \!\! dz\, f^2(\vx)
\end{equation}
has to be used.  As shown before $B_0$ (e.g. by Eq.~\ref{eq:B2first}),
and $\lambda_B$ (e.g. by Eq.~\ref{eq:lB}) can be derived for a given
window function $f(\vx)$ using $C_{\rm RM}^\obs(\vrp)$. For a good
choice of the window function, one gets
\begin{equation}
\label{eq:chi2}
\chi^2_{\rm av} = \frac{1}{A_{\Omega}}\, \int\!dx_\perp^2\,
\chi^2(\vxp) \approx 1\,,
\end{equation}
and larger values if the true and assumed models differ
significantly.  The model discriminating power lies also in the
spatial distribution $\chi^2(x_\perp)$, and not only in its
global average. If some large scale trends are apparent, e.g. that
$\chi^2(x_\perp)$ is systematically higher in more central or more
peripheral regions of the Faraday screen, then such a model for
$f(\vx)$ should be disfavoured. This can be tested by averaging
$\chi^2(x_\perp)$ e.g. in radial bins for a roughly spherical screen,
as a relaxed galaxy cluster should be, and checking for apparent
trends.

We note, that this method of model testing can be regarded as a
refined Laing-Garrington effect \citep{1988Natur.331..149L,
1988Natur.331..147G}: The more distant radio cocoon of a radio galaxy
in a galaxy cluster is usually more depolarised than the nearer radio
cocoon due to the statistically larger Faraday depth. This is observed
whenever the observational resolution is not able to resolve the RM
structures. Here, we assume that the observational resolution is
sufficient to resolve the RM structures, so that a different depth of
some part of the radio source observed, or a different average
magnetic energy profile leads to a different statistical Faraday depth
$\langle \RM(\vxp)^2 \rangle$. Since this can be tested by suitable
statistics, e.g. the simple $\chi^2$ statistic proposed here,
incorrect models can be identified.

It may be hard in an individual case to disentangle the effect of
changing the total depth $z_{\rm s}$ of the used polarised radio
source if it is embedded in the Faraday screen, and the effect of
changing $B_0^{\rm obs}$, since these two parameters can be quite
degenerate. However, there may be situations in which the geometry is
sufficiently constrained because of additional knowledge of the source
position, or statistical arguments can be used if a sufficiently large
sample of similar systems were observed.

\section{Fourier space formulation\label{sec:Fspace}}
\subsection{Basics\label{sec:basicsFSF}}

We use the following convention for the Fourier
transformation of a $n$-dimensional function $F(\vx)$:
\begin{eqnarray}
\hat{F}(\vk) &=& \int \!\! d^nx \, F(\vx) \,e^{i\,\vk \vec{\cdot}
\vx}\\ 
F(\vx) &=& \frac{1}{(2\,\pi)^n} \int \!\! d^nk \, \hat{F}(\vk)
\,e^{-i\,\vk \vec{\cdot} \vx}.
\end{eqnarray}

The Fourier transformed isotropic magnetic autocorrelation tensor
reads
\begin{equation}
\label{eq:Mfs}
\hat{M}_{ij}(\vk) =  \hat{M}_N(k) \,\left(\delta_{ij} -
\frac{k_i\,k_j}{k^2}\right) - i \eps_{ijm}\, \frac{k_m}{k}\, \hat{H}(k)\,,
\end{equation}
where we have directly used the $\vec{\nabla}\vec{\cdot}\vec{B} = 0$
condition in the form $k_i\,\hat{M}_{ij}({\vk}) = 0$ to reduce the
degrees of freedom to two components, a normal and helical part. The
two corresponding spherically symmetric functions in $k$-space are given
in terms of their real space counterparts as:
\begin{eqnarray}
 \hat{M}_N(k)  \!\! &=& \!\! \int \!\! d^3r M_N(r)\,e^{i\,\vk\vec{\cdot}\vr}
 = \!\! 4\pi \!\! \int_0^\infty \!\!\!\!\!\!\!\! dr\,r^2 M_N(r)
\frac{\sin(k\, r)}{k\,r}\\
 \hat{H}(k) \!\! &=& \!\! \frac{d}{dk} \hat{M}_H(k) \!\! = \!\!
 \frac{d}{dk} \! \int \!\! d^3r M_H(r)\,e^{i\,\vk\vec{\cdot}\vr},
 \nonumber\\
  \!\! &=& \!\! \frac{4\pi}{k} \!\!
 \int_0^\infty \!\!\!\!\!\! dr\, r^2 M_H(r) \,\frac{k\, r\,\cos(k\, r) -
 \sin(k\, r)}{k\,r}.
\end{eqnarray}
One can also introduce the Fourier transformed trace of the
autocorrelation tensor $\hat{w}(\vk) = \hat{M}_{ii}(\vk) =
2\,\hat{M}_N(k)$. A comparison with the transformed $zz$-component of
the autocorrelation tensor 
\begin{equation}
\hat{M}_{zz} (\vk) = \hat{M}_N(k) \left( 1 - k_z^2/k^2 \right)
\end{equation}
reveals that in the $k_z = 0$ plane these two functions are identical
(up to a constant factor $2$). Since the 2-d Fourier transformed
normalised RM map is also identical to this, as a transformation of
Eq.~\ref{eq:Cperp1st} shows, we can state
\begin{equation}
\label{eq:FTedCperp}
\hat{C}_\perp(\vkp) =\hat{M}_{zz} (\vkp,0) = \frac{1}{2} \, \hat{w}(\vkp,0).
\end{equation}
This Fourier-space version of Eq.~\ref{eq:Cproj} says, that the 2-d
transformed RM map reveals the $k_z = 0$ plane of $\hat{M}_{zz}
(\vk)$, which in the isotropic case is all what is required to
reconstruct the full magnetic autocorrelation $\hat{w}(k) =
2\,\hat{C}_\perp(k)$.

\subsection{Power spectra\label{sec:power}}

\begin{figure*}[t]
\psfig{figure=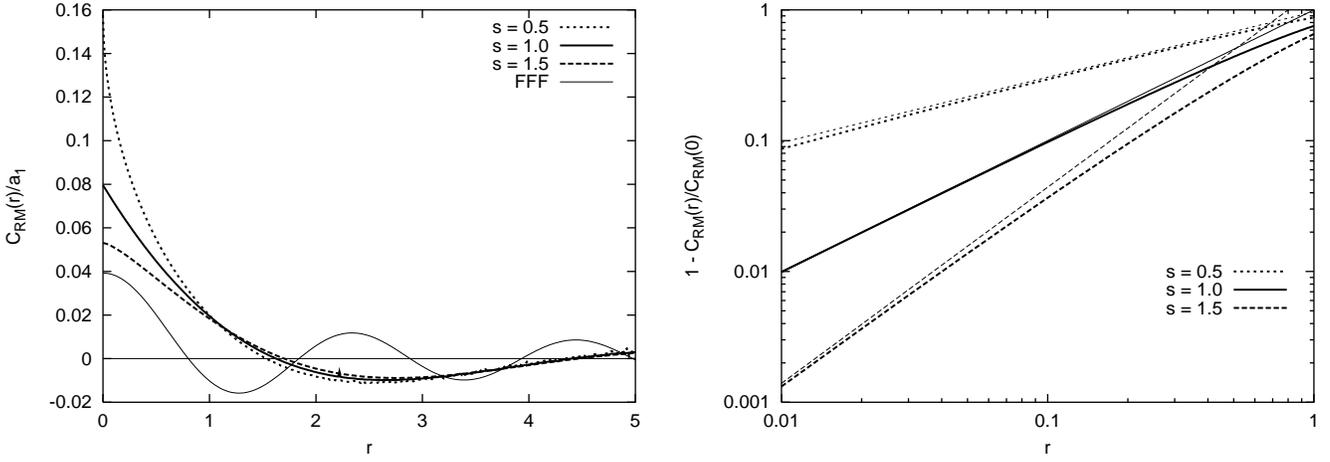,width= \textwidth,angle=0}
\caption[]{\label{fig:crm} Left: $C_{\perp}(r_\perp)$ for power-law
spectra of magnetic fluctuations $\hat{w}(k) = \hat{w}_0 \, k^{-s-2}\,
\vec{1}_{\{k_1<k<k_2\}}$ within the range $k_1 = 1$ to $k_2 = 10^4$
with normalisation $\hat{w}_0 = 1$ and different slopes $s$ as
labelled. For comparison, the thin line shows the RM correlation
function expected for a single-scale power spectrum, as it can arise
for isotropic linear force-free fields ($C_\perp(r_\perp) = \pi
\,\langle B^2 \rangle \, {\rm J}_0(k_{\rm F}\,r_\perp)/(2\,k_{\rm F})$
with $k_{\rm F} = 3$, and $\langle B^2 \rangle = 0.075$; see
Sect. \ref{sec:fff}).
Right: $1-C_{\perp}(r_\perp)/C_{\perp}(0)$ in logarithmic units for
the same power-law spectra.  The thin lines are the asymptotic spectra
given by the first two terms in Eq.~\ref{eq:CperpApprox}.}
\end{figure*}

We recall that the power spectrum $P_{[F]}(\vk)$ of a function
$F(\vx)$ is given by the absolute-square of its Fourier transformation
$P_{[F]}(\vk) = |\hat{F}(\vk)|^2$.  The WKT states that the Fourier
transformation of an autocorrelation function $C_{[F]}(\vr)$,
estimated within a window with volume $V_n$ (as in
Eq.~\ref{eq:CRMobs}), gives the (windowed) power spectrum of this
function, and vice versa:
\begin{equation}
P_{[F]} (\vk) = V_n\, \hat{C}_{[F]}(\vk)\,.
\end{equation} 

The WKT allows us to write the Fourier transformed autocorrelation
tensor as
\begin{equation}
\label{eq:defMinFS}
\hat{M}_{ij}(\vk) = \frac{1}{V} \langle \hat{B}_i(\vk)\,\overline{\hat{B}_j(\vk)}
\rangle\,,
\end{equation}
where $V$ denotes the volume of the window function, which is for
practical work with RM maps often the probed effective volume $V=
V_{[f]}$ as defined in Sect. \ref{sec:defVf}.

Thus, the 3-d magnetic power spectrum (the Fourier transformed
magnetic autocorrelation function $w(\vr)$) can be directly connected
to the one-dimensional magnetic energy spectrum in the case of
isotropic turbulence:
\begin{equation}
\label{eq:epsB}
\eps_B(k) \,dk = \frac{4\,\pi\,k^2}{(2\pi)^3}\,\frac{\hat{w}(k)}{8
\,\pi}\,dk = \frac{k^2\,\hat{w}(k)}{2\,(2\pi)^3}\,dk\,,
\end{equation}
where we wrote $\hat{w}(\vk) =\hat{w}(k)$ due to isotropy. The WKT
also connects the 2-dimensional Fourier-transformed RM map with the
Fourier transformed RM autocorrelation function:
\begin{equation}
\label{eq:CperpA}
\hat{C}_\perp(k_\perp) = \frac{\langle |\hat{\RM}(k_\perp)|^2
\rangle}{a_1\,A_{\Omega}} .
\end{equation}

Thus, by comparing Eqs.  \ref{eq:FTedCperp}, \ref{eq:epsB}, and
\ref{eq:CperpA} one finds that the magnetic energy spectrum is most
easily measured from a given observation by simply Fourier
transforming the map $\RM(\vxp)$, and averaging this over rings in
$\vkp$-space:
\begin{equation}
\label{eq:epsBm}
\eps_B^\obs(k) = \frac{k^2}{a_1\,A_{\Omega}\,(2\pi)^4}
\int_0^{2\,\pi}\!\!\!\!\!\! d\phi \, |\hat{\RM}(\vkp)|^2
\end{equation}
where $\vkp = k \,(\cos{\phi},\sin{\phi})$. Eq.~\ref{eq:epsBm} gives a
direct model independent observational route to measure the turbulent energy
spectrum. The average magnetic energy density can be easily obtained from
this via
\begin{equation}
\label{eq:measureBinFS}
\eps_B^\obs = \int_0^{\infty}\!\!\!\!\!\! dk \,\eps_B^{\rm
 obs}(k) = \int \!\! d^2 k_\perp\, \frac{ k_\perp\,
 |\hat{\RM}(\vkp)|^2}{a_1\,A_{\Omega}\,(2\pi)^4}\,,
\end{equation}
where the last integration extends over the Fourier transformed RM map
and can be done in practice by summing over pixels.

Also the correlation lengths can be expressed in terms of $\hat{w}(k)$:
\begin{equation}
\label{eq:lambdainFS}
\lambda_B = \pi \frac{\int_{0}^{\infty} dk \,k\,
\hat{w}(k)}{\int_{0}^{\infty} dk \,k^2\, \hat{w}(k)},
\;\;
\lambda_{\rm RM} = 2 \frac{\int_{0}^{\infty} dk \,
\hat{w}(k)}{\int_{0}^{\infty} dk \,k\, \hat{w}(k)}.
\end{equation}
Thus, the RM correlation length has a much larger weight on the
large-scale fluctuations than the magnetic correlation length
has. Equating these two length-scales, as sometimes done in the
literature, is at least questionable in the likely case of a broader
turbulence spectrum. In typical situations (e.g. for a broad maximum
of the magnetic power spectrum as often found in hydrodynamical
turbulence) one expects $\lambda_B < \lambda_{\rm RM}$. Since the
former is the one which enters the magnetic field estimates by using
the measured RM-dispersion,
\begin{equation}
\label{eq:Btraditional}
\langle B^2 \rangle = \frac{2\,C_{\rm RM}(0)}{a_1\,\lambda_B} =
\frac{2\, \langle \RM^2 \rangle}{a_0^2\,n_{\rm e}^2\,L\,\lambda_B},\,
\mbox{with}\, L = \frac{V_{[f]}}{A_\Omega},
\end{equation}
using the easily measurable $\lambda_{\rm RM}$ instead of $\lambda_B$
likely underestimates the magnetic field strength.

The isotropic magnetic autocorrelation function can be expressed
as
\begin{equation}
w(r) = \frac{4\,\pi}{(2\,\pi)^3} \int_0^\infty\!\!\!\!\!\! dk\,k^2\,
\hat{w}(k) \frac{\sin(k\,r)}{k\,r}.
\end{equation}
Similarly, the RM autocorrelation function can be written as
\begin{equation}
C_\perp(r_\perp) = \frac{1}{4\,\pi}\,\int_0^\infty\!\!\!\!\!\! dk\,k\,\hat{w}(k)\,{\rm J}_0(k\,r_\perp)\,,
\end{equation}
where ${\rm J}_n(x)$ is the n-th Bessel function. In order to analyse the
behaviour of the RM autocorrelations close to the origin, it is useful
to rewrite the last equation as:
\begin{equation}
\label{eq:Cperpsplit1}
C_\perp(r_\perp) = \int_0^\infty\!\!\!\!\!\!
dk\,\frac{k\,\hat{w}(k)}{4\,\pi} - \int_0^\infty\!\!\!\!\!\!
dk\,\frac{k\,\hat{w}(k)}{4\,\pi}\,(1-{\rm J}_0(k\,r_\perp)).
\end{equation}
The first term gives $C_\perp(0)$, and the second describes how
$C_\perp(r_\perp)$ approaches zero for $r_\perp \rightarrow \infty$.

\subsection{Power-law power spectra\label{sec:plps}}

In many cases the small-scale magnetic energy spectrum is a power-law,
say $\eps_B(k) = \eps_0 k^{-s}$ (e.g. $s=5/3$ for Kolmogorov-like
turbulence, as expected if the magnetic fields were shaped by a mostly
hydrodynamical turbulence) or $w_B(k) = \hat{w}_0 k^{-2-s}$ for
$k_1<k<k_2$ (with $k_1 \ll k_2$). For the behaviour of
$C_\perp(r_\perp)$ on such scales Eq.~\ref{eq:Cperpsplit1} can be
written as
\begin{equation}
\label{eq:CperpApprox}
C_\perp(r_\perp) = C_\perp(0) - G(s)\,\hat{w}_0\,r_\perp^s + R_{[w]}(r_\perp)\,, 
\end{equation}
where
\begin{eqnarray}
C_\perp(0) &=& \int_0^\infty\!\!\!\!\!\!
dk\,\frac{k\,\hat{w}(k)}{4\,\pi},\\
G(s) &=& \frac{\Gamma(\frac{2-s}{2})}{2^{s+2}\,\pi\,s\,
\Gamma(\frac{2+s}{2})}\;\mbox{for}\; 0<s<2,\; \mbox{and}\\
R_{[w]}(r) &=& \frac{1}{4\,\pi}\! \int_0^\infty\!\!\!\!\!\! dk\,k\,(\hat{w}(k) - \hat{w}_0\,
k^{-s-2})\,(1-{\rm J}_0(k\,r)).
\end{eqnarray}
On small scales (more specifically for $k_1 \ll 1/r_\perp \ll k_2$)
and for well behaved power spectra outside $k_1 \ll k \ll k_2$ the
term $R_{[w]}(r)$ is negligibly small. We therefore propose to fit
\begin{equation}
\label{eq:CpPowerLaw} 
C_{\rm RM}(r_\perp) \approx C_0 - C_1 \, r_\perp^s
\end{equation}
to the inner part of an observationally determined RM correlation
function. From this, the turbulence spectral index $s$, $C_\perp(0) =
C_0/a_1$, and the power-law normalisation $w_0 = C_1/(a_1\,G(s))$ can
be inferred. A rough estimate of $k_1$ and $k_2$, the scales on which
the spectrum deviates from the power-law, can also be obtained from
finding the $r_\perp$-values, where the fit becomes poor. A more
accurate determination of these scales can always be done in the
Fourier domain (see Eqs.~\ref{eq:FTedCperp}, \ref{eq:CperpA} and
\ref{eq:epsBm}). We therefore recommend the usage of
Eq.~\ref{eq:CpPowerLaw} more for consistency checks and rapid and rough
{\it fit-by-eye} diagnostics of the steepness of the magnetic power
spectrum rather than for high-precession analysis.

The shape of the RM correlation function close to the origin allows a
direct read-off of the type of magnetic turbulence. A top-down
scenario, where most of the energy resides on large scales ($s>1$) and
the smaller scales are populated by a turbulent cascade as in the
Kolmogorov-, Kraichnan-, and Goldreich-Sridhar-phenomenologies, leads
to a flat cusp at the origin, and a convex shape near to it. A
bottom-up magnetic turbulence scenario, where the fields originate on
small scales and are enlarged by shear flows or other inverse cascade
actions ($s<1$), leads to a sharp cusp at the origin, and a concave
slope next to it. A spectral energy distribution with as much energy
on small as on large scales ($s= 1$) leads to a linear cusp at the
origin. The behaviour of $C_\perp(r_\perp)$ for these three cases is
illustrated in Fig.~\ref{fig:crm}.

\subsection{Helical correlations \& force-free fields \label{sec:fff}}

Faraday rotation maps do not contain information about the helical
part of the autocorrelation tensor. Therefore, additional information
is required in order to be able to measure the helical
correlations. For example any relation between the helical and
non-helical components would be sufficient.

In order to give an example for such additional information, we
discuss the case of force-free fields (FFFs). The condition for FFFs
reads
$
\vec{\nabla} \vec{\times}\vec{B} = k_{\rm F}\,\vec{B}\,,
$ where $k_{\rm F}$ can in general be a function of position. For
simplicity, we restrict $k_{\rm F}$ to be spatially constant. Such
so-called {\it linear FFFs} lead to a very simple structure of the
components of $\hat{M}_{ij}(\vk)$: For $k \not = k_{\rm F}$ all
components vanish, and for $k = k_{\rm F}$ one gets $\hat{M}_N(k_{\rm
F}) = - \hat{H}(k_{\rm F})$, leaving the magnetic energy density (or
the helicity) as the only remaining free parameter for a given
characteristic wave-vector $k_{\rm F}$. FFFs are therefore also called
{\it maximally helical fields}.

From the fact that for a linear FFF only one spherical shell in
wave-vector space is populated with magnetic power, the spatial
autocorrelation function is easily obtained as 
\begin{equation}
w_{\rm F}(r) = \langle B^2 \rangle \, \sin(k_{\rm F}\,r)/(k_{\rm F}\,r)
\end{equation}
 \citep{1999PhRvL..83.2957S} and the RM
autocorrelation as 
\begin{equation}
C_\perp(r_\perp) = \pi \,\langle B^2 \rangle \, {\rm J}_0(k_{\rm
F}\,r_\perp)/(2\,k_{\rm F}).
\end{equation}
This function is shown on the left side of Fig.~\ref{fig:crm}.

\subsection{Finite window functions\label{sec:windowcorr}}

Here, we discuss the effect of a finite window function on magnetic
field estimates, in order to possibly correct for the bias made with
the robust weighting scheme introduced in Sect. \ref{sec:rm}. Taking a
finite window function $f(\vr)$ into account, Eq.~\ref{eq:FTedCperp}
becomes
\begin{equation}
\label{eq:CperpwithWindow}
\hat{C}_\perp^\obs (\vkp) = \frac{1}{2}\, \int \!\!\! d^3q\, 
\hat{w}(q)\, \frac{q_\perp^2}{q^2}\,\,
W(\vkp-\vec{q})
\end{equation}
where we introduced
\begin{equation}
W(\vk) =  \frac{|\hat{f}(\vk)|^2}{{(2 \pi)^3}\,V_{[f]}}\,,
\end{equation}
and used the identity $\hat{M}_{zz}(\vec{k}) =
\frac{1}{2}\hat{w}(k)\,(1-k_z^2/k^2)$. Without the term
${q_\perp^2}/{q^2}$ the convolution integral in
Eq.~\ref{eq:CperpwithWindow} would describe a redistribution of the
magnetic power within Fourier space, which conserves the total
magnetic energy. But the term ${q_\perp^2}/{q^2}$ leads to some loss
of magnetic power.

In any situation in which there is substantial magnetic power on
scales comparable or larger than the window size
Eq.~\ref{eq:CperpwithWindow} can be used to estimate the response of
the observation to the magnetic power on a given scale $p$ by
inserting $\hat{w}(q) = \delta(q-p)$. Ideally, this is then used
within a matched-filter analysis or as the response matrix in a
maximum-likelihood reconstruction of the underlying power-spectra. The
computation of the response matrix relating input power $\hat{w}(q)$
and measured signal $\hat{C}_\perp^\obs (\vkp)$ can be cumbersome
since in general a 2- or 3-dimensional integral (depending if one uses
the delta function) has to be evaluated for each matrix element.
Therefore we restrict our discussion here to three highly symmetric,
idealised cases, and an approximative treatment of a more realistic
configuration, which should give a feeling for the general
behaviour.

\begin{figure}[t]
\psfig{figure=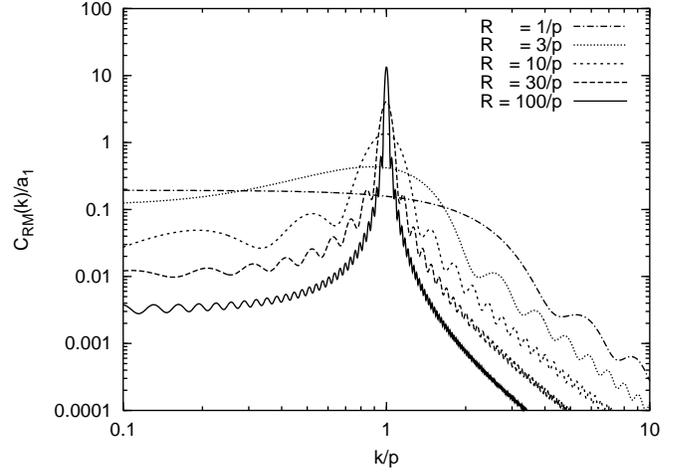,width= 0.5\textwidth,angle=0}
\caption[]{\label{fig:Cperp2} Response in $\hat{C}_\perp(k_\perp)$ to
a single-scale magnetic power signal $\hat{w}(k) = \delta(k-p)$ at
wavenumber $p$ for different radii $R$ of a cylindrical window function. }
\end{figure}

\
{\bf A cylindrical window:} Suppose a circular radio source with
radius $R$ is seen through a very deep Faraday screen, so that the
depth $L_z$ can be approximated to be infinite long. The window
function $f(\vec{x})= \vec{1}_{\{x_\perp <R\}}$ leads to 
\begin{equation}
W(\vk) =\frac{{\rm J}_1^2(k_\perp\,R)}{\pi\,k_\perp^2}\, \delta(\vk_z),
\end{equation}
where $\delta(k)$ is the Dirac's delta function. Inserting this into 
Eq.~\ref{eq:CperpwithWindow} gives
\begin{equation}
\hat{C}_\perp^\obs (\vkp) = \frac{p}{2\,\pi} \int_0^{2\pi}
\!\!\!\!\!\!\!\! d\phi\;\, \frac{{\rm J}_1^2(\sqrt{k_\perp^2 + p^2 - 2
\,p\,k_\perp \cos\phi }\,R)}{k_\perp^2 + p^2 - 2
\,p\,k_\perp \cos\phi},
\end{equation}
which is shown in Fig.~\ref{fig:Cperp2}. One can clearly see that with
increasing window size the response becomes more and more
delta-function-like.

\begin{figure}[t]
\psfig{figure=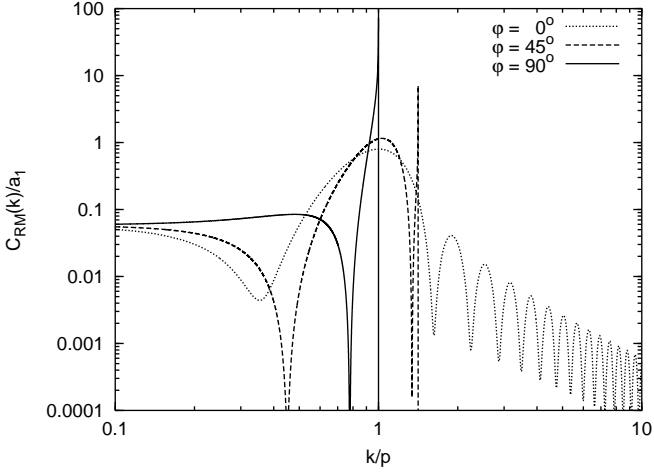,width= 0.5\textwidth,angle=0}
\caption[]{\label{fig:Cperp3} Response in
$\hat{C}_\perp(k_\perp\,(\cos\phi, \sin\phi))$ to
a single-scale magnetic power signal $\hat{w}(k) = \delta(k-p)$ at
wavenumber $p$ for a sheet-like window seen edge-on with diameter
$L_x = 10/p$ for different wavevector orientations $\phi$ with respect to the
sheet normal.}
\end{figure}

{\bf A sheet-like window seen edge-on:} Suppose a very elongated radio
source is seen through a deep Faraday screen, so that the window
function is approximated by $f(\vx) = \vec{1}_{\{0<x<L_x\}}$. This
gives
\begin{equation}
W(\vk) = \Delta_{L_x}(k_x)\, \delta(k_y)\, \delta(k_z),
\end{equation}
where we introduced a shortcut for the Fourier transformed
1-d-box-window of size $L$:
\begin{equation}
\Delta_{L}(k) =
\frac{L}{2\pi}\, \frac{\sin^2(k\,L/2)}{(k\,L/2)^2} \rightarrow
\delta (k) \;\mbox{for}\; L \rightarrow \infty\,.
\end{equation}
This leads to a response of
\begin{equation}
\hat{C}_\perp^\obs (\vkp) = 
\vec{1}_{\{k_y<p\}}\, \frac{p}{2\, t}\, (\Delta_{L_x}(k_x-t) +\Delta_{L_x}(k_x+t))
\end{equation}
where we wrote $t = (p^2-k_y^2)^{1/2}$ for brevity. The anisotropic
response of this window is much more delta-function like for
wavevectors oriented along the sheet axis than oriented perpendicular
to it, as can be seen in Fig.~\ref{fig:Cperp3}.

\begin{figure}[t]
\psfig{figure=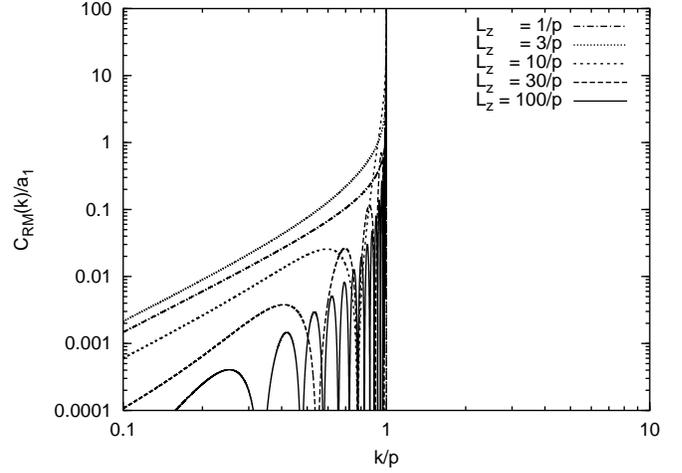,width= 0.5\textwidth,angle=0}
\caption[]{\label{fig:Cperp1} Response in $\hat{C}_\perp(k_\perp)$ to
a single-scale magnetic power signal $\hat{w}(k) = \delta(k-p)$ at
wavenumber $p$ for a sheet-like window seen face-on for different diameters
$L_z$.}
\end{figure}

{\bf A sheet-like window seen face-on:} Suppose the window is an
infinitely extended homogeneous layer of thickness $L_z$ in $z$-direction,
e.g. a magnetised skin layer of a large radio source, so that $f(\vx)
= \vec{1}_{\{0 <z <L_z\}}$, and
\begin{equation}
W(\vk) = \Delta_{L_z}(k_z)\, \delta^2(\vkp),
\end{equation}
where $\delta^2(\vkp)$ is the 2-d Dirac's delta function.
The response function 
\begin{equation}
\hat{C}_\perp^\obs (\vkp) = 
\vec{1}_{\{k_\perp<p\}}\, \frac{k_\perp^2}{p\,\sqrt{p^2-k_\perp^2}}
\,\Delta_{L_z}(\sqrt{p^2 - k_\perp^2})
\end{equation}
is isotropic, and is shown in Fig.~\ref{fig:Cperp1}. In this geometry,
power gets scattered out of the $k_z =0$ plane due to the finite
window size in $z$-direction. This leads to a loss of magnetic power
in any measurement, which does not correct for this bias.

The observed magnetic energy density estimated with the help of
Eq.~\ref{eq:measureBinFS} can be shown to be related to the real
magnetic power spectrum via
\begin{equation}
\eps_B^\obs = \int_0^\infty \!\!\!\!\! dk\; \eps_B(k) \,D(k)\,,
\end{equation}
where the weighting function
\begin{eqnarray}
D(k) &=& \frac{3}{2\,k\,L_z} \left[ {_1{\rm F}_2\! \left(\frac{1}{2};\,
\frac{1}{2}, 2;\, -\left(\frac{k\,L_z}{2}\right)^2 \right) -1} \right]
\\
&\approx& \frac{k\,L_z}{\sqrt{(16/3)^2 + (k\,L_z)^2}} = \left\{ 
\begin{array}{lll}
3\,k\,L_z/16 &;& k\,L_z \ll 1\\
1 &;& k\,L_z \gg 1
\end{array}
\right.
\end{eqnarray}
describes the relative contribution of different parts of the
spectrum. Note that $\eps_B(k) \,D(k)$ is not the observed power
spectrum, but only the contribution to the magnetic energy estimate, that
$_a{\rm F}_b$ denotes the hypergeometric function, and that the
asymptotic approximation has an overall accuracy of better than
$5\%$. Since $D(k) < 1$ everywhere, it is obvious that the derived
magnetic fields are underestimated, especially if there is substantial
magnetic power on scales comparable and larger than the window size.

In principle, it is possible to correct for any bias, if the
window function is reliably known and if the statistical sampling is
sufficiently good even on the large scales so that dividing the
observed magnetic power spectrum by the weighting function gives
sensible results (and not just amplifies the noise).

{\bf Approximative treatment of a realistic window:} In a realistic
situation, often a relatively small sized radio galaxy is seen through
a deep Faraday screen. In such a case the depth can again be
approximated to be infinite for the purpose of the Fourier-space
window:
\begin{equation}
W(\vk) = \delta(k_z) \,W_\perp(\vkp)\,,
\end{equation}
where we introduced the projected Fourier-window
\begin{equation}
W_\perp (\vkp) = \frac{| \hat{f}_\perp(\vkp)|^2}{(2\, \pi)^2\,A_{[f_\perp]}}\,,
\end{equation}
which results from a projected window function:
\begin{equation}
f_\perp^2(\vxp) = \int \!\!\! dz\, f^2(\vx)/L_z\;\mbox{with}\; A_{[f_\perp]} = \int
\!\!\! d^2x_\perp \, f_\perp^2(\vxp)\,.
\end{equation}
Here, $L_z$ is an arbitrary but fixed reference length, e.g. the
typical source size $L_z = V_{[f]}/A_\Omega$.  In the case that the
observationally measured magnetic power spectrum results from a
spherical average over the data
\begin{equation}
\hat{w}^\obs(k) = \frac{1}{2\,\pi}\,\int_0^{2\pi} \!\!\!\!\!\!\!\! d\phi\, 
2\,\hat{C}_\perp^\obs(k\,(\cos\phi, \sin\phi))
\end{equation}
the response to a delta function-like magnetic power-spectrum $\hat{w}(q) =
\delta(q-p)$  is given by
\begin{eqnarray}
\hat{w}_p^\obs (k_\perp) &=& \frac{2\,p}{\pi} \int \!\!\! d^2 q_\perp\,
\frac{W_\perp(\vec{q}_\perp )\, \vec{1}_{\{|k_\perp - p| \le q_\perp
\le k_\perp + p\} }}{\sqrt{4\,q_\perp^2 \,p^2 - (q_\perp^2 + p^2 -
k_\perp^2)^2}}
\\
&=& \frac{2\,p}{\pi}\, \int_{|k_\perp - p|}^{k_\perp + p}
\!\!\!\!\!\!\!\!\!\!
 dq \,\,\,\,
\frac{ q\, \int_0^{2\pi} \!\!\!\!\!\!\! d\phi\, W_\perp (q (\cos\phi,
\sin\phi))}{\sqrt{4\,q^2 \,p^2 - (q^2 + p^2 - k_\perp^2)^2}}.
\end{eqnarray}
This approximative response can be easily computed numerically for
any model window function. In many cases it should be sufficiently
accurate to estimate the effect of a finite observational window on
the derived magnetic power spectrum.  As a consistency check, we
verified that the limit of an infinitely extended radio source, which
can be written as $u\rightarrow 0$ in $W_\perp(\vec{q}_\perp) =
\vec{1}_{\{q_\perp \le u \}}/(\pi\, u^2)$, leads to $\hat{w}_p^\obs(q) =
\delta(q-p)$ as it should.

\subsection{Testing isotropy\label{sec:tstIso}}

Since isotropy of the magnetic field statistics is a crucial
ingredient of the proposed analysis, it is important to test if
indications of anisotropy are present, and to see how anisotropy can
affect the results.

Anisotropy can manifest itself in two different ways: (a) The Fourier
space magnetic power distribution can be anisotropic, by being
not only a function of $k$ but a full function of $\vk$, and (b) the
magnetic power tensor itself can be anisotropic. Certainly both
flavours of anisotropy can be present simultaneously. However, their
effects can be well separated, so that we discuss them one by one.

Before doing so, we note that the relation
\begin{equation}
\label{eq:C-Mzz}
\hat{C}_\perp(\vkp) = \hat{M}_{zz} (\vkp,0)
\end{equation}
is completely independent on assumptions on isotropy. The RM power
spectrum still reveals the $k_z = 0$ plane of the $zz$-component of
the magnetic power tensor. The condition of isotropy had allowed to
use the measured information as a representative probe of (a) the full
Fourier-space and (b) the other diagonal elements of the magnetic
tensor.

{\bf (a) Anisotropic power spectrum}: Eq. \ref{eq:C-Mzz} shows that an
anisotropic power spectrum can be detected, since it leads very likely 
to an anisotropic RM power map if the anisotropy is not aligned with
the $z$-direction by chance. Since the latter can not be excluded, it
is hard to prove isotropy. On the other hand, a perfect alignment of
the line-of-sight and the anisotropy axis is not very likely. By
studying a number of independent Faraday screens, an anisotropic power
spectrum can be ruled out on a statistical basis. Furthermore, by
co-adding the signals of several systems, statistical isotropy can be
enforced, even if an individual system is anisotropic. However, in
order to be able to co-add different observations, the window
functions have to be well understood. Especially the scaling of the
average magnetic field energy density with location within the Faraday
screen and from screen-to-screen should be known. Since this
is still poorly known, it is worth to check if indications of
anisotropy are present in every dataset itself.

A good way to check for indications of anisotropy is by eye inspection
of maps of $\hat{C}_\perp(\vkp)$ or equivalently $C_\perp(\vrp)$ or by
comparing profiles which were calculated using different angular
slices.  A more quantitative estimate of apparent anisotropy can be
obtained by the use of multipole moments. Since the dipole moment
vanishes due to mirror symmetries in $\hat{C}_\perp(\vkp)$ and
$C_\perp(\vrp)$, the first non-trivial multipole is the quadrupole
moment:
\begin{equation}
Q_{ij}^{(\gamma)} = \int\! d^2k_\perp \, \hat{C}_\perp(\vkp)\, (2\,k_i\,k_j -
k_\perp^2\,\delta_{ij})\,k_\perp^{-\gamma}
\end{equation}
(and similarly for $C_\perp(r_\perp)$). Here we allowed for a
weighting factor $k_\perp^{-\gamma}$ in order to balance the
contributions from different scales.  $Q^{(\gamma)}$This should be
compared to the second moment of the (weighted) distribution
\begin{equation}
P_{2}^{(\gamma)} = \int\! d^2k_\perp \, \hat{C}_\perp(\vkp)\, k_\perp^{2-\gamma} \;\;\;
\;\;(\mbox{and similarly for}\,\, C_\perp(r_\perp)),
\end{equation}
e.g. by calculating the ratio $R^{(\gamma)}=(|Q_{xx}^{(\gamma)}| +
|Q_{xy}^{(\gamma)}|)/(2\,P_2^{(\gamma)})$. In an isotropic case this
number should be close to zero, in a strongly anisotropic case it can
become comparable with one. We suggest to apply this test to the
real-space and the Fourier-space data, since in the last case isotropy
is mostly tested on large spatial scales, and in the second case on
small spatial scale. Attention has to be given to the fact that the
real-space quadrupole moment is sensitive (for small $\gamma$) to any
non-circularity of the window function which can affect large
$\vrp$. The Fourier-space quadrupole moment can be affected (for small
$\gamma$) by any ellipticity of the synthesised beams of the
observations since this manifests itself at large $\vkp$. Therefore
the integration range for the quadrupole moments and the second
moments may be better restricted to intermediate radii in real- and in
Fourier-space.

{\bf (b) Anisotropic tensor}: In the anisotropic case, the only
constraint on the magnetic autocorrelation tensor is
$k_i\,\hat{M}_{ij}(\vk) = 0$ due to the divergence-freeness of the
magnetic fields. This translates for the $z$-components into $k_x \,
\hat{M}_{xz} (\vk) + k_y \, \hat{M}_{yz} (\vk) + k_z \, \hat{M}_{zz}
(\vk) =0$, which leaves the observable $\hat{M}_{zz} (\vkp,0)$
absolutely unconstrained since $k_z = 0$. $\hat{M}_{zz}$ can therefore
be an arbitrary function of $\vkp$. However, if it is not circularly
symmetric, this can be detected with the methods described above. 

In order to have a working example of an anisotropic part
$\hat{M}^{a}_{ij}$ of the magnetic tensor we assume that a preferred
direction $\vec{a}(k)$ exists, so that 
\begin{equation}
\hat{M}^{a}_{ij}(\vk) = \left(a_i - \frac{k_i\,a_l\,k_l}{k^2}\right)
\left(a_j - \frac{k_j\,a_l\,k_l}{k^2}\right)\,.
\end{equation}
This is an intrinsically anisotropic tensor, which fulfils
$k_i\,\hat{M}_{ij}^a(\vk) = 0$. The Faraday-observable component is
$\hat{M}^{a}_{ij}(\vkp,0) = a^2_z(k_\perp)$. Its influence on the RM
statistics can not be discriminated from an isotropic contribution if
$a(k)$ depends only on $|\vk|$. The assumption of isotropy would
therefore lead to an incorrect estimate of the field strength, since
the measured $zz$-component is assumed to be representative for all
components. However, if the signal from a number of similarly
anisotropic Faraday screens are co-added the errors compensate
statistically, if no correlation of the anisotropic direction and the
line-of-sight are present.

Furthermore, since any anisotropy of the magnetic power tensor should
have a physical cause, e.g. a large-scale shear flow in the Faraday
active medium, an accompanying anisotropic power spectrum is very
likely, which can principally be detected by the methods described
above (a).

Finally, if anisotropy turns out to be inherently present in Faraday
screens, one might replace Eqs. \ref{eq:Mreal} and \ref{eq:Mfs} by a
more complex, anisotropic model in order to be able to extract
information from individual screens. In that case this work
may help as a guideline for such a more elaborate analysis.

\section{Magnetic structures\label{sec:magStr}}
\subsection{Autocorrelation\label{sec:struct}}
The possibility exists that the magnetic fields of a Faraday screen
consist of several distinct magnetic structures like flux ropes,
magnetic tori etc. If the positions and orientations of the structures
can be regarded as statistically independent the magnetic
autocorrelation function can be written as
\begin{equation}
w(r) = \sum_{s} n_{s} \, W_{\rm s}(r)\, ,
\end{equation}
where all types $s$ of structures present with space
density $n_{s}$ are summed up. A structure $s$ with field configuration
$\vec{B}_s(\vx)$ has an intrinsic isotropically averaged
(unnormalised) autocorrelation function
\begin{equation}
W_{s}(r) = \frac{1}{4\,\pi}\, \int \!\! d^2 \Omega \,\int \!\!  d^3x \,
\vec{B}_s(\vx)  \vec{\cdot}  \vec{B}_s(\vx+ \vec{\Omega}\, r)\,, 
\end{equation}
where the first integration covers the unit sphere. 

For a magnetic structure, which consists of a mostly constant magnetic
field $B_s$ within the volume $V_s$, and negligible field strength
elsewhere, the autocorrelation function is asymptotically for small
$r$
\begin{equation}
W_{s}(r) = B_s^2\, V_s \,( 1 - r/l_s )\,,
\end{equation}
where $l_s$ is a typical length-scale of the structure, roughly given
by $l_s \sim V_s/A_s$ with $A_s$ the surface area of the
structure. If only a single type of structure is present, we get
asymptotically
\begin{equation}
\label{eq:wStructempl}
w(r) = B_s^2\, \eta_B \,( 1 - r/l_s )\,,
\end{equation}
where $\eta_B = n_s \,V_s$ is the magnetic volume filling factor.

In order to calculate the RM autocorrelation of such a Faraday screen,
we use as a toy model $w(r)$ from Eq.~\ref{eq:wStructempl} as long as
$r<r_{\rm max}$, and otherwise $w(r)=0$. Eq.~\ref{eq:ACvol} would then
requires $r_{\rm max} = \frac{4}{3} \,l_s$, but the actual choice is
only important for numerical values of constants of proportionality,
and not for the qualitative shape of the RM autocorrelation function
at the origin. Integrating Eq.~\ref{eq:Cproj} leads to an asymptotic
expansion of the form
\begin{equation}
\label{eq:logcusp}
C_\perp(r_\perp) = C_0 - \,\left[C_1 + C_2 \ln(\frac{l_s}{r_\perp} )
\right]\, \left(\frac{r_\perp}{l_s} \right)^2\,,  
\end{equation}
which gives a flat central slope, nearly a parabola, but still
having a tiny logarithmic cusp. The constants $C_0$, $C_1$, and
$C_2$ depend on the details of the outer slope of $w(r)$, e.g. on the
ratio $r_{\rm max}/l_s$, so that their numerical values are model
dependent.

We summarise that a Faraday screen built from structural elements with
internally constant magnetic fields, and only a single characteristic
length-scale leads to a flat central autocorrelation function, with at
most a logarithmic cusp of the form given by Eq.~\ref{eq:logcusp}.

\subsection{Filling factor\label{sec:fillingF}}

Although there exist characteristic shapes of the RM autocorrelation
function $C_\perp(r_\perp)$ in the case of a patchy magnetised Faraday
screen, as demonstrated in Sect.\ref{sec:struct}, the presence of such
patches can not be deduced from $C_\perp(r_\perp)$ alone. Since the phase
information is missing, the special form of the cusp arising from
magnetic structures as given by Eq.~\ref{eq:logcusp} can not be
distinguished from a complete random phase turbulence with steep
power-law like spectra with spectral index $s \approx 2$, as can be
seen from comparison with Eq.~\ref{eq:CpPowerLaw}.

In order to measure the patchiness of the magnetic field distribution
in galaxy clusters \cite{2001ApJ...547L.111C} used the area filling
factor $\eta_\RM$ of the line-of-sight of extended radio sources which
do not show any RM due to the Faraday screen. For their sources, they
concluded that $\eta_\RM > 95\%$.

If the magnetic fields are in flux-rope like structures, with typical
length $l_\|$ and diameter $l_\perp$, the cross section of a flux-rope
to be intersected by a line-of-sight is of the order $l_\| \,
l_\perp$. Their volume filling factor is $\eta_{B} \approx l_\| \,
l_\perp^2\, n_{\rm rope}$. If their locations can be regarded as
being uncorrelated, the number $K$ of flux ropes intersected by a
line-of-sight of length $L_{\rm los}$ is Poisson-distributed: $P(K) =
\Lambda^K \, \exp(-\Lambda)/(K!)$ with $\Lambda \approx \eta_{B}\,
L_{\rm los}/l_\|$. From that it follows by inserting $K=0$ that
\begin{equation}
\eta_{B} \approx \frac{l_\|}{L_{\rm los}}\, \ln(
\frac{1}{1-\eta_\RM}). 
\end{equation}

For filaments of length $l_\| \approx 10$ kpc and lines-of-sight of
$L_{\rm los} \approx$ 500 kpc only a small subvolume $\eta_{B} > 0.05$
of the clusters actually needs to be magnetised in order to give the
large area filling factor found by \cite{2001ApJ...547L.111C}.

Another constraint for the magnetic filling factor can be obtained
from energetic arguments. The magnetic field energy density in
magnetised regions can be expected to be below the environmental
thermal energy density $\eps_{\rm th}$, since otherwise a magnetic
structure would expand until it reaches pressure equilibrium. Since
the autocorrelation analysis of RM maps is able to provide the volume
averaged magnetic field energy density $\langle \eps_B \rangle$, the magnetic
volume filling factor can be constrained to be
\begin{equation}
\eta_B > \frac{\langle \eps_B \rangle}{\eps_{\rm th}} = 0.8\cdot 10^{-2}\,
\frac{\langle B^2 \rangle}{\mu{\rm G}^2} \, \left[ \frac{n_{\rm e}}{10^{-3}
\, {\rm cm^{-3}}} \, \frac{kT}{{\rm keV}} \right]^{-1}.
\end{equation}
In cases of relatively strong average magnetic energy densities, as
e.g. in cooling flow clusters, or in case of physical arguments
requiring a much lower than equipartition field strength, this can
give a tight constraint on the volume filling factor.

\section{Observational artefacts\label{sec:artefacts}}
\subsection{Beam smearing}

The finite size $l_{\rm beam}$ of a synthesised beam of a radio
interferometer should smear out RM structures below the beam size, and
therefore can lead to a smooth behaviour of the measured RM
autocorrelation function at the origin, even if the true
autocorrelation function has a cusp there. Substantial changes of the
RM on the scale of the beam can lead to beam depolarisation, due to
the differentially rotated polarisation vectors within the beam area
\citep{1985A&A...146..392C, 1988Natur.331..149L,
1988Natur.331..147G}. Since beam depolarisation is in principal
detectable by its frequency dependence, the presence of sub-beam
structure can be noticed, even if not resolved
\citep{1991MNRAS.250..726T, 1998ApJ...505..921M}. The magnetic power
spectrum derived from a beam smeared RM map should cut-off at large $k
\sim \pi/l_{\rm beam}$.

\subsection{Noise}
Instrumental noise can be correlated on several scales, since radio
interferometers sample the sky in Fourier space, where each antenna
pair baseline measures a different $k_\perp$-vector. It is difficult
to understand to which extend noise on a telescope antenna baseline
pair will produce correlated noise in the RM map, since several
independent polarisation maps at different frequencies are combined in
the map making process.  We therefore discuss only the case of
spatially uncorrelated noise, as it may result from a pixel-by-pixel
RM fitting routine. This adds to the RM autocorrelation function
\begin{equation}
C_{\rm noise}^\obs(\vxp, \vrp) = \sigma_{\RM,\rm noise}^2(\vxp) \,\delta^2(\vrp)\,.
\end{equation}
In Fourier space, this leads to a constant error for $\hat{w}(k)$
\begin{equation}
\hat{w}_{\rm noise}(k) = 2\, \langle \sigma_{\RM,\rm noise}^2\rangle
\end{equation}
and therefore to an artificial component in the magnetic power
spectrum $\eps_B^\obs(k)$ increasing by $k^2$.

If it turns out that for an RM map with an inhomogeneous noise map (if
provided by an RM map construction software) the noise affects the
small-scale power spectrum too severely, one can try to reduce this by
down-weighting noisy regions with a suitable choice of the data
weighting function $h(\vxp)$ which was introduced in
Sect. \ref{sec:rm} for this purpose.

\subsection{RM steps due to the $n\pi$-ambiguity}

An RM map is often derived by fitting the wavelength-square behaviour
of the measured polarisation angles. Since the polarisation angle is
only determined up to an ambiguity of $n\pi$ (where $n$ is an
integer), there is the risk of getting a fitted RM value which is
off by $m\,\Delta \RM$ from the true one. $m$ is an integer,
and $\Delta\RM = \pi\,(\lambda_{\rm min}^{2} - \lambda_{\rm
max}^{2})^{-1}$ is a constant depending on the used wavelength range from
$\lambda_{\rm min}$ to $\lambda_{\rm max}$.

This can lead to artifical jumps in RM maps, which will affect the RM
autocorrelation function and therefore any derived magnetic power
spectrum. In order to get a feeling for this we model the possible
error by an additional component in the derived RM map:
\begin{equation}
\RM^{\rm amb}(\vxp) = \sum_i\, m_i\; \Delta\RM\; \vec{1}_{\{ \vxp \in
\Omega_i \}}\,,
\end{equation}
where $\Omega_i$ is the area of the $i$-th RM patch, and $m_i$ is an
integer, mostly +1 or -1. Assuming that different patches are
uncorrelated, the measured RM autocorrelation function is changed by
an additional component, which should be asymptotically for small
$r_\perp$
\begin{equation}
C_{\RM}^{\rm amb}(r_\perp) = \Delta\RM^2 \, \eta_{\rm amb}\, \left( 1 -
\frac{r_\perp}{l_{\rm amb}} \right),
\end{equation}
where $\eta_{\rm amb}$ is the area filling factor of the ambiguity
patches in the RM map, and $l_{\rm amb}$ a typical patch diameter.

Comparing this with Eq.~\ref{eq:CpPowerLaw} shows that the artificial
power induced by the $n\pi$-ambiguity mimics a turbulence energy
spectrum with slope $s=1$, which would have equal power on all
scales. A steep magnetic power spectrum can therefore possibly be
masked by such artifacts.

Fortunately, for a given observation the value of $\Delta\RM$ is
known and one can search an RM map for the occurrence of steps by
$\Delta\RM$ over a short distance (not necessarily one pixel) in
order to detect such artifacts.

\section{Conclusions\label{sec:concl}}

We have investigated the statistics of Faraday rotation maps on the
level of the autocorrelation function and the power spectrum. We
proposed ways to analyse extended Faraday maps in order to reconstruct
the magnetic autocorrelation tensor (Eqs.~\ref{eq:defM} and
\ref{eq:defMinFS}) from which quantities like the average field
strength, the magnetic energy spectrum, and their autocorrelation
length can be obtained (Eq.~\ref{eq:lB}). We showed that under the
assumption of isotropy of the observed magnetic field ensemble the
symmetric part of the magnetic autocorrelation tensor
(Eqs.~\ref{eq:Mreal} and \ref{eq:Mfs}) can be reconstructed. This
makes use of the condition $\vec{\nabla} \vec{\cdot} \vec{B} = 0$ and
the additional assumption (which can be tested a-posteriori) that the
gradient scale of the electron density (e.g. the core radius of a
galaxy cluster) is much larger than the typical field
length-scale. The anti-symmetric or helical part of the magnetic
correlation tensor can only be measured if additional information is
available, e.g. in the case of force-free fields
(Sect. \ref{sec:fff}).

The assumption of isotropy of the magnetic field statistics should be
justified in cases where a sufficiently large volume of the screen is
probed. In principle, it can also be tested by searching for
non-circular distortions of the 2-dimensional autocorrelation function
(Sect. \ref{sec:tstIso}).

A further test for statistical isotropy and sufficient sampling of the
field fluctuations is the fact that if these conditions are given in a
finite Faraday screen (which cannot maintain infinitely long
correlations) the rotation measure (RM) autocorrelation area
(Eq.~\ref{eq:ACarea}) has to vanish. This means that there is a
balance between the positively and negatively valued areas of the
autocorrelation function.  In practice, one would require it to be
much smaller than the RM autocorrelation length squared. We note
that e.g. the popular magnetic cell-model, in which cells are filled
by from cell-to-cell independently oriented and internally homogeneous
magnetic fields, does not have these properties, since it violates the
required $\vec{\nabla} \vec{\cdot} \vec{B} = 0$ condition.

Our approach is meant to be applied directly to real data. Effects of
incomplete information, due to the limited extent of polarised radio
sources, are properly treated in form of a window function
(Eqs.~\ref{eq:window}, \ref{eq:exptCRMobs}, and
\ref{eq:CperpwithWindow}). This window function contains additional
information on the screen geometry, and allows for proper bookkeeping
of data weighting, in case of noisy data being analysed. Since the
window function also requires some working hypothesis about the
average magnetic energy density profile of the Faraday screen, ways to
test it a posteriori are sketched (Sect. \ref{sec:wtst}).

The most efficient way to analyse Faraday rotation maps leads through
Fourier space (Sect. \ref{sec:Fspace}). The Fourier transformation of
a map gives direct insight into the magnetic energy spectrum
(Eqs.~\ref{eq:epsB} and \ref{eq:epsBm}), which fully specifies the
magnetic autocorrelation function (Eq.~\ref{eq:FTedCperp}). Many
quantities of interest can be obtained from it, such as the average field
strength (Eq.~\ref{eq:measureBinFS}), the correlation lengths
(Eq.~\ref{eq:lambdainFS}), and the bias resulting from the used window
function (Sect. \ref{sec:windowcorr}).

The Fourier domain formulation gives also important insight into the
real space behaviour of the RM autocorrelation function
(Sect. \ref{sec:plps}): A power-law magnetic energy spectrum leads to
a cusp at the origin of this function, where the shape of the cusp is
determined by the power law index $s$. A steep spectrum $s>1$ (as
e.g. expected for turbulent cascades) leads to a flat cusp, whereas a
flat spectrum $s<1$ gives a pronounced sharp cusp. The limiting case
$s=1$ with equal power on all scales leads to a linear (decreasing)
behaviour of the RM autocorrelation function close to the origin.

Such cusps of the RM autocorrelation (or power-law spectra in Fourier
space) can be signatures of turbulent cascades, but they can also
occur in other situations. We demonstrated that a Faraday screen
which is composed of finite magnetic structures of roughly
constant field strength lead to a flat cusp, too
(Sect. \ref{sec:struct}). We show ways to constrain the magnetic
volume filling factor for such magnetic field models
(Sect. \ref{sec:fillingF}).  Observational artifacts, like noise or
jumps in the measured RM values due to the so called $n\pi$-ambiguity,
are able to produce sharp cusps (Sect. \ref{sec:artefacts}). We
therefore stress the importance to check maps for such distortions and
describe ways to do this.

A very important result of this work is that the magnetic
autocorrelation length $\lambda_B$ (Eq.~\ref{eq:lB}) is in general not
identical to the autocorrelation length of the RM fluctuations
$\lambda_\RM$ (Eq.~\ref{eq:lRM}). $\lambda_\RM$ is much more strongly
weighted towards the large-scale part of the magnetic power spectrum
than $\lambda_B$ (Eq.~\ref{eq:lambdainFS}). In typical astrophysical
situations a broad spectral energy distribution can be expected so
that $\lambda_B$ will be much smaller than $\lambda_\RM$. Since
$\lambda_B$ enters the classical formulae to estimate magnetic field
strength from Faraday measurements $\langle B^2\rangle \propto \langle
\RM^2 \rangle / \lambda_B$ (see Eq.~\ref{eq:Btraditional}), but
$\lambda_\RM$ is sometimes used instead, we expect that several
published magnetic field estimates from Faraday rotation maps are
actually underestimates.

We hope that our work aids and stimulates further observational and
theoretical work on the exciting field of Faraday rotation
measurements of cosmic magnetic fields in order to give us deeper
insight in their fascinating origins and roles in the Universe.

\begin{acknowledgements}
We thank Kandaswamy Subramanian for many discussions and important
suggestions. We acknowledge further stimulating discussions on RM maps
with Matthias Bartelmann, Tracy E. Clarke, Klaus Dolag, Luigina
Feretti, Gabriele Giovannini, Federica Govoni, Philipp P. Kronberg,
and Robert Laing. We thank Matthias Bartelmann and Tracy Clarke for
comments on the manuscript.  TAE thanks for the hospitality at the
Istituto di Radioastronomia at CRN in Bologna in April 2000, where
several of the discussions took place.  This work was done in the
framework of the EC Research and Training Network {\it The Physics of
the Intergalactic Medium}.

\end{acknowledgements}



\begin{thebibliography}{}

\bibitem[\protect\astroncite{{Beck}}{2001}]{2001SSRv...99..243B}
{Beck}, R., 2001,
\newblock {Space Science Reviews} {99}, 243

\bibitem[\protect\astroncite{{Beck} et~al.}{1996}]{1996ARA&A..34..155B}
{Beck}, R., {Brandenburg}, A., {Moss}, D., {Shukurov}, A., {Sokoloff}, D.,
  1996,
\newblock {\araa} {34}, 155

\bibitem[\protect\astroncite{{Bicknell} et~al.}{1990}]{1990ApJ...357..373B}
{Bicknell}, G.~V., {Cameron}, R.~A., {Gingold}, R.~A., 1990,
\newblock {\apj} {357}, 373

\bibitem[\protect\astroncite{{Carilli} \&
  {Taylor}}{2002}]{2002ARA&A..40..319C}
{Carilli}, C.~L., {Taylor}, G.~B., 2002,
\newblock {\araa} {40}, 319

\bibitem[\protect\astroncite{{Cho} et~al.}{2002}]{2002ApJ...564..291C}
{Cho}, J., {Lazarian}, A., {Vishniac}, E.~T., 2002,
\newblock {\apj} {564}, 291

\bibitem[\protect\astroncite{{Clarke} et~al.}{2001}]{2001ApJ...547L.111C}
{Clarke}, T.~E., {Kronberg}, P.~P., {B{\" o}hringer}, H., 2001,
\newblock {\apjl} {547}, L111

\bibitem[\protect\astroncite{{Conway} \& {Strom}}{1985}]{1985A&A...146..392C}
{Conway}, R.~G., {Strom}, R.~G., 1985,
\newblock {\aap} {146}, 392

\bibitem[\protect\astroncite{{Dolag} et~al.}{1999}]{1999A&A...348..351D}
{Dolag}, K., {Bartelmann}, M., {Lesch}, H., 1999,
\newblock {\aap} {348}, 351

\bibitem[\protect\astroncite{{Dolag} et~al.}{2002}]{2002A&A...387..383D}
{Dolag}, K., {Bartelmann}, M., {Lesch}, H., 2002,
\newblock {\aap} {387}, 383

\bibitem[\protect\astroncite{{Dolag} et~al.}{2001}]{2001A&A...378..777D}
{Dolag}, K., {Schindler}, S., {Govoni}, F., {Feretti}, L., 2001,
\newblock {\aap} {378}, 777

\bibitem[\protect\astroncite{{Dreher} et~al.}{1987}]{1987ApJ...316..611D}
{Dreher}, J.~W., {Carilli}, C.~L., {Perley}, R.~A., 1987,
\newblock {\apj} {316}, 611

\bibitem[\protect\astroncite{{Eilek} \& {Owen}}{2002}]{2002ApJ...567..202E}
{Eilek}, J.~A., {Owen}, F.~N., 2002,
\newblock {\apj} {567}, 202

\bibitem[\protect\astroncite{{En{\ss}lin} \&
  {Biermann}}{1998}]{1998AA...330...90E}
{En{\ss}lin}, T.~A., {Biermann}, P.~L., 1998,
\newblock {\aap} {330}, 90

\bibitem[\protect\astroncite{{Feretti}}{1999}]{1999dtrp.conf....3F}
{Feretti}, L., 1999,
\newblock in P.~S. H.~{B{\"o}hringer}, L.~{Feretti} (ed.), {Diffuse Thermal and
  Relativistic Plasma in Galaxy Clusters}, Vol. 271 of {MPE-report}, p.~3

\bibitem[\protect\astroncite{{Garrington} et~al.}{1988}]{1988Natur.331..147G}
{Garrington}, S.~T., {Leahy}, J.~P., {Conway}, R.~G., {Laing}, R.~A., 1988,
\newblock {\nat} {331}, 147

\bibitem[\protect\astroncite{{Goldreich} \&
  {Sridhar}}{1997}]{1997ApJ...485..680G}
{Goldreich}, P., {Sridhar}, S., 1997,
\newblock {\apj} {485}, 680

\bibitem[\protect\astroncite{{Govoni} et~al.}{2002}]{astro-ph/0211292}
{Govoni}, F., {Feretti}, L., {Murgia}, M., {Taylor}, G., {Giovannini}, G., and
  {Dallacasa}, D., 2002,
\newblock in {Matter and Energy in Clusters of Galaxies}, {Bowyer}, S. and
  {Hwang}, C.-Y., Chung-Li, Taiwan,
\newblock astro-ph/0211292

\bibitem[\protect\astroncite{{Grasso} \&
  {Rubinstein}}{2001}]{2001PhR...348..163G}
{Grasso}, D., {Rubinstein}, H.~R., 2001,
\newblock {\physrep} {348}, 163

\bibitem[\protect\astroncite{{Kolatt}}{1998}]{1998ApJ...495..564K}
{Kolatt}, T., 1998,
\newblock {\apj} {495}, 564

\bibitem[\protect\astroncite{{Kosowsky} \& {Loeb}}{1996}]{1996ApJ...469....1K}
{Kosowsky}, A., {Loeb}, A., 1996,
\newblock {\apj} {469}, 1

\bibitem[\protect\astroncite{{Kronberg}}{1994}]{1994RPPh...57..325K}
{Kronberg}, P.~P., 1994,
\newblock {Reports of Progress in Physics} {57}, 325

\bibitem[\protect\astroncite{{Kulsrud}}{1999}]{1999ARA&A..37...37K}
{Kulsrud}, R.~M., 1999,
\newblock {\araa} {37}, 37

\bibitem[\protect\astroncite{{Laing}}{1988}]{1988Natur.331..149L}
{Laing}, R.~A., 1988,
\newblock {\nat} {331}, 149

\bibitem[\protect\astroncite{{Maron} \&
  {Goldreich}}{2001}]{2001ApJ...554.1175M}
{Maron}, J., {Goldreich}, P., 2001,
\newblock {\apj} {554}, 1175

\bibitem[\protect\astroncite{{Melrose} \&
  {Macquart}}{1998}]{1998ApJ...505..921M}
{Melrose}, D.~B., {Macquart}, J.-P., 1998,
\newblock {\apj} {505}, 921

\bibitem[\protect\astroncite{{Ohno} et~al.}{2002}]{ohno2002}
{Ohno}, H., {Takada}, M., {Dolag}, K., {Bartelmann}, M., {Sugiyama}, N.,
  2002,
\newblock {\apj} in press,
\newblock arXiv:astro-ph/0206278

\bibitem[\protect\astroncite{{Oren} \& {Wolfe}}{1995}]{1995ApJ...445..624O}
{Oren}, A.~L., {Wolfe}, A.~M., 1995,
\newblock {\apj} {445}, 624

\bibitem[\protect\astroncite{{Rees}}{1987}]{1987QJRAS..28..197R}
{Rees}, M.~J., 1987,
\newblock {\qjras} {28}, 197

\bibitem[\protect\astroncite{{Sridhar} \&
  {Goldreich}}{1994}]{1994ApJ...432..612S}
{Sridhar}, S., {Goldreich}, P., 1994,
\newblock {\apj} {432}, 612

\bibitem[\protect\astroncite{{Subramanian}}{1999}]{1999PhRvL..83.2957S}
{Subramanian}, K., 1999,
\newblock {Physical Review Letters} {83}, 2957

\bibitem[\protect\astroncite{{Taylor} et~al.}{2001}]{2001MNRAS.326....2T}
{Taylor}, G.~B., {Govoni}, F., {Allen}, S.~W., {Fabian}, A.~C., 2001,
\newblock {\mnras} {326}, 2

\bibitem[\protect\astroncite{{Taylor} \& {Perley}}{1993}]{1993ApJ...416..554T}
{Taylor}, G.~B., {Perley}, R.~A., 1993,
\newblock {\apj} {416}, 554

\bibitem[\protect\astroncite{{Tribble}}{1991}]{1991MNRAS.250..726T}
{Tribble}, P.~C., 1991,
\newblock {\mnras} {250}, 726

\bibitem[\protect\astroncite{{Widrow}}{2002}]{2002RvMP...74..775W}
{Widrow}, L.~M., 2002,
\newblock {Reviews of Modern Physics} {74}, 775

\bibitem[\protect\astroncite{{Wielebinski} \&
  {Krause}}{1993}]{1993A&ARv...4..449W}
{Wielebinski}, R., {Krause}, F., 1993,
\newblock {\aapr} {4}, 449

\bibitem[\protect\astroncite{Willson}{1970}]{1970MNRAS.151....1W}
Willson, M. A.~G., 1970,
\newblock {\mnras} {151}, 1

\end{thebibliography}

\end{document}